\documentclass[preprint,10pt]{elsarticle}
\pdfoutput=1


\makeatletter
\def\ps@pprintTitle{%
   \let\@oddhead\@empty
   \let\@evenhead\@empty
   \def\@oddfoot{\reset@font\hfil\thepage\hfil}
   \let\@evenfoot\@oddfoot
}
\makeatother

\usepackage{titlesec}
\usepackage{xcolor}
\usepackage{enumitem}
\usepackage[hidelinks]{hyperref}
\hypersetup{colorlinks=true,urlcolor=black}
\usepackage[T1]{fontenc}\usepackage{lmodern}

\usepackage{amsmath,amsfonts,amssymb,mathtools}
\usepackage{caption}
\usepackage{subcaption}
\usepackage{booktabs}
\usepackage{placeins}
\usepackage{tikz}
\usepackage{pgfplots}
\pgfplotsset{compat=newest}
\usepgfplotslibrary{colorbrewer}
\usetikzlibrary{pgfplots.colorbrewer}
\usetikzlibrary{patterns}

\usepackage{multirow}
\usepackage{adjustbox}
\usepackage{afterpage}

\usepackage[utf8]{inputenc}

\definecolor{rev_red}{RGB}{227,26,28}%{217,95,2}
\definecolor{rev_green}{RGB}{51,160,44}%{102,166,30}
\definecolor{rev_blue}{RGB}{31,120,180}%{231,41,138}
\definecolor{rev_orange}{RGB}{255,127,0}
\definecolor{rev_purple}{RGB}{106,61,154}
\newcommand{\reviewerOne}[1]{#1}
\newcommand{\reviewerTwo}[1]{#1}
\newcommand{\reviewerThree}[1]{#1}
\newcommand{\reviewerOneThree}[1]{#1}
\newcommand{\reviewerTwoThree}[1]{#1}
\usepackage[margin=1in]{geometry}
\biboptions{sort&compress}
\begin{document}

\begin{frontmatter}

\title{Nonintrusive Manufactured Solutions for Non-Decomposing Ablation in Two Dimensions}

\author[freno]{Brian A.~Freno}
\ead{bafreno@sandia.gov}
\author[freno]{Brian R.~Carnes}
\author[brunini]{Victor E.~Brunini}
\author[freno]{Neil R.~Matula}

\address[freno]{Sandia National Laboratories, Albuquerque, New Mexico 87185, USA}
\address[brunini]{Sandia National Laboratories, Livermore, CA 94551, USA}

\begin{abstract}
\reviewerOne{Code verification is a necessary} step towards establishing credibility in computational physics simulations.  \reviewerOne{It is used to assess the correctness of the implementation of the numerical methods within the code, and it is a continuous part of code development.  Code verification} is typically performed using exact and manufactured solutions.  However, exact solutions are \reviewerOne{often} limited, and manufactured solutions generally require the invasive introduction of an artificial forcing term within the source code, such that the code solves a modified problem for which the solution is known.  
\reviewerThree{The equations for some physics phenomena, such as non-decomposing ablation, yield infinite analytic solutions, but the boundary conditions may eliminate these possibilities.}
%For some physics phenomena, such as non-decomposing ablation, there are many possible exact solutions to the governing equations, but the boundary conditions may render these impractical and eliminate such conveniences as separation of variables.  
%
For such phenomena, however, we can manufacture the terms that comprise the boundary conditions to obtain exact solutions.  In this paper, we present a nonintrusive method for manufacturing solutions for non-decomposing ablation in two dimensions, which does not require the addition of a source term.
\end{abstract}


\end{frontmatter}

%\clearpage
%==============================================================================
\section{Introduction} %=======================================================
%==============================================================================

Ablation plays an important role in many scientific and engineering applications, including fire protection, surgical procedures, combustion, and manufacturing.  An understanding of ablative processes is particularly critical in the realm of hypersonic flight, where the ablating material serves to carry heat energy away from the vehicle and its payload.  Since weight and cost are paramount concerns for flight vehicles, accurate prediction of the rate of mass and energy removal is essential, as it allows the designer to minimize heat shield weight under the constraint of maintaining adequate thermal protection.  Furthermore, the spatial distribution of the ablation process directly affects the aerodynamic performance of the vehicle.  Hence, credible predictions of ablative processes are crucial for the design of safe and efficient high-speed vehicles.

% Code Verification

As with the computational simulation of any physical phenomenon, it is necessary to assess the implementation and the suitability of the underlying models in order to develop confidence in the simulation results.  These assessments typically fall into two complementary categories: verification and validation.  Validation evaluates the appropriateness of the models instantiated in the code for representing the relevant physical phenomena, and is typically performed through comparison with experimental data.  Verification, on the other hand, assesses the correctness of the numerical solutions produced by the code, through comparison with the expected theoretical behavior of the implemented numerical methods.  Following Roache~\cite{roache_1998},  Salari and Knupp~\cite{salari_2000}, and Oberkampf and Roy~\cite{oberkampf_2010}, verification can be further divided into the activities of code verification and solution verification.  Solution verification involves the estimation of the numerical error for a particular simulation, whereas code verification assesses the correctness of the implementation of the numerical methods within the code.  A review of code and solution verification is presented by Roy~\cite{roy_2005}.

This paper focuses on code verification.  The discretization of the governing equations of the physical models necessarily incurs a truncation error, and the solution to the discretized equations therefore incurs an associated discretization error.  In the most basic sense of verification, if the discretization error tends to zero as the discretization is refined, the consistency of the code is verified~\cite{roache_1998}.  This may be taken a step further by examining not only consistency, but the rate at which the error decreases as the discretization is refined.  The code may then be verified by comparing this rate to the expected theoretical order of accuracy of the discretization scheme.  Unfortunately, this approach requires knowledge of the exact solution to the problem at hand, and exact solutions to problems of engineering interest are rare.  Hence, manufactured solutions are frequently employed to produce problems of sufficient complexity with known solutions~\cite{roache_2001}.

% Literature Review
% Verification has been performed on computational physics codes associated with several fluid dynamics phenomena, including 
% laminar flows~\cite{roy_2004}, 
% turbulent flows~\cite{bond_2007,veluri_2010,oliver_2012,eca_2016,ghebali_2017,wang_2021},
% flows with finite Knudsen numbers~\cite{myong_2019},
% viscous flows in shock tubes~\cite{zhou_2018},
% hypersonic reacting flows~\cite{freno_2021},
% dense gas-particle flows~\cite{meng_2019},
% fluid--structure interaction~\cite{etienne_2012}, 
% heat transfer in fluid--solid interaction~\cite{veeraragavan_2016}, 
% multiphase flows~\cite{brady_2012}, 
% and 
% radiation hydrodynamics~\cite{mcclarren_2008},
% as well as on the discretization of the gradient operator for finite volume methods~\cite{syrakos_2017}. 
\reviewerThree{Code verification is necessary to gather credibility evidence during the development of any simulation code that solves discretized equations.  Examples of code verification have been demonstrated for} 
%Code verification has been performed on
computational physics codes associated with several physics disciplines, including fluid dynamics~\cite{roy_2004,bond_2007,veluri_2010,oliver_2012,eca_2016,freno_2021}, solid mechanics~\cite{chamberland_2010}, fluid--structure interaction~\cite{etienne_2012}, heat transfer in fluid--solid interaction~\cite{veeraragavan_2016}, multiphase flows~\cite{brady_2012,lovato_2021}, radiation hydrodynamics~\cite{mcclarren_2008}, electrodynamics~\cite{ellis_2009}, and electromagnetism~\cite{marchand_2013,freno_em_mms_2020,freno_em_mms_quad_2021}.
Code-verification techniques for ablation have been presented by Hogan et al.~\cite{hogan_1994}, Blackwell and Hogan~\cite{blackwell_1994}, and Amar et al.~\cite{amar_2008,amar_2009,amar_2011} for simple exact solutions.  Additionally, a manufactured solution for heat conduction has been presented in Amar et al.~\cite{amar_2011}.
A nonintrusive approach to manufactured solutions for non-decomposing ablation in one dimension has been introduced by Freno et al.~\cite{freno_ablation}.  

\reviewerTwo{%
For non-decomposing ablation, thermal decomposition is considered negligible.  The governing equation is the heat equation with insulated boundaries on the non-ablating surfaces and a heat flux on the ablating surface that arises from convection, energy loss from ablation, and radiation. 
}

In this paper, we extend the approach of Reference~\cite{freno_ablation} to introduce a nonintrusive manufactured solutions approach for non-decomposing ablation in two dimensions, which introduces additional considerations, such as nontrivial mesh deformation and coordinate system choice.  These nonintrusive manufactured solutions avoid the need to modify the code to introduce a forcing term.  This property is particularly favorable for cases where the code is either proprietary or has other characteristics that make it inaccessible to one performing code verification.  The trade-off is reduced freedom compared to traditional manufactured solutions; however, this reduction does not limit the ability to exercise the capabilities of the code.

\reviewerTwoThree{%
This approach is designed for equations with complicated boundary conditions.  For non-decomposing ablation and other physics phenomena with governing equations that permit analytic solutions, this approach avoids the need to modify the code to introduce a forcing term.  However, for other phenomena, such as decomposing ablation, analytic solutions require too many limiting assumptions, and a forcing term needs to be introduced.  For these phenomena, after manufacturing the solution, the parameters can be manufactured to satisfy the boundary conditions.
}

% In this paper, we present a nonintrusive approach to manufacturing solutions for ablation.  The ablation code relies heavily on external data, which serve as problem parameters.  These parameters include the material properties (e.g., specific heat capacity and thermal conductivity), as well as the heating conditions (e.g., char ablation rate, heat transfer coefficient, wall enthalpy, and recovery enthalpy).

\reviewerTwoThree{%
For non-decomposing ablation, we} begin this process by optionally transforming the governing equations and deriving solutions to them.  These solutions satisfy the boundary conditions on the non-ablating surfaces but not on the ablating surface.  With these solutions, we manufacture the remaining parameters to satisfy the boundary condition on the ablating surface.  
Like traditional manufactured solutions, certain desirable properties of the underlying functions, such as a sufficient number of finite nontrivial derivatives and elementary function composition, take precedence over being physically realizable.
Through this approach we can modify external data, rather than modifying the code to introduce a forcing term.

% Outline
This paper is organized as follows.  
Section~\ref{sec:governing_equations} describes the heat equation, as well as the ablation contribution and the domain evolution.  
Section~\ref{sec:mms} details our approach for verifying the accuracy of the discretization.  
Section~\ref{sec:solutions} provides derivations of exact solutions to the heat equation to account for ablation in Cartesian and polar coordinates.
Section~\ref{sec:bc} describes how the ablation parameters are manufactured to satisfy the boundary conditions.
Section~\ref{sec:results} demonstrates this methodology with numerical examples.
Section~\ref{sec:conclusions} summarizes this work.

\section{Governing Equations}
\label{sec:governing_equations}

For a solid, the energy equation due to heat conduction is
\begin{align}
\frac{\partial}{\partial t}(\rho e) + \nabla\cdot \mathbf{q} = 0.
\label{eq:energy}
\end{align}
The specific internal energy $e$ can be modeled by $e=e_0+\int_{T_0}^T c_p(\hat{T}) d\hat{T}$, where $c_p=c_p(T)$ is the specific heat capacity, and the heat flux $\mathbf{q}$ can be modeled by Fourier's law, 
\begin{align}
\mathbf{q}=-k(T)\nabla T,
\label{eq:fourier}
\end{align}
where $k(T)$ is the thermal conductivity of the isotropic material.

If the material density $\rho$ is constant, %invariant with respect to time and temperature, 
\eqref{eq:energy} becomes
\begin{align}
\rho c_p(T)\frac{\partial T}{\partial t}  - \nabla\cdot\left(k(T)\nabla T\right) = 0.
\label{eq:energy_2}
\end{align}
When the material properties are constants, such that $k=\bar{k}$ and $c_p=\bar{c}_p$, \eqref{eq:energy_2} reduces to the constant-coefficient heat equation,
%
%\begin{align*}
$\frac{\partial T}{\partial t}   - \bar{\alpha}\Delta T = 0$,
%\label{eq:energy_const_prop}
%\end{align*}
%
with thermal diffusivity $\bar{\alpha}=\frac{\bar{k}}{\reviewerTwo{\rho} \bar{c}_p}$.

%======================================================================
\subsection{Ablation and Boundary Conditions} %===========================
%======================================================================
\label{sec:ablation_bc}

\begin{figure}%[!b]
\centering
\begin{tikzpicture}
%\useasboundingbox (-7,-4) (7,4);
\path (-4.5,0) -- (8.5,0);
\def\H{6}
\def\W{3}
\def\shiftleft{-2.5}
\def\shiftright{5}
\def\tbar{5}
\def\nscale{.7}
%\draw (-\W/2,-\H/2) -- (\W/2,-\H/2) -- (\W/2,\H/2) -- (-\W/2,\H/2) -- (-\W/2,-\H/2);
\def\nincr{4}
\foreach \i in {0,...,\nincr}%
{
\def\timefrac{\i/\nincr}
\def\maxblack{40}
\def\minblack{100}
\pgfmathsetmacro{\colorfrac}{\i/\nincr*\minblack + (\nincr-\i)/\nincr*\maxblack}

% Cartesian domain at each time
\draw[color=black!\colorfrac,domain=0:1, smooth, variable=\xi] plot ({\shiftleft+\W/2 -\W/4*\timefrac*(1+2*sin(90*\xi))},{\xi*\H-\H/2}) -- (\shiftleft-\W/2,\H/2) -- (\shiftleft-\W/2,-\H/2) -- cycle;

% Cartesian arrows at each time
\draw[color=black!\colorfrac, draw opacity=0,domain=0:.2, smooth, variable=\xi,->,>=stealth] plot ({\shiftleft+\W/2 -\W/4*\timefrac*(1+2*sin(90*\xi))},{\xi*\H-\H/2});
\draw[color=black!\colorfrac, draw opacity=0,domain=0:.4, smooth, variable=\xi,->,>=stealth] plot ({\shiftleft+\W/2 -\W/4*\timefrac*(1+2*sin(90*\xi))},{\xi*\H-\H/2});
\draw[color=black!\colorfrac, draw opacity=0,domain=0:.6, smooth, variable=\xi,->,>=stealth] plot ({\shiftleft+\W/2 -\W/4*\timefrac*(1+2*sin(90*\xi))},{\xi*\H-\H/2});
\draw[color=black!\colorfrac, draw opacity=0,domain=0:.8, smooth, variable=\xi,->,>=stealth] plot ({\shiftleft+\W/2 -\W/4*\timefrac*(1+2*sin(90*\xi))},{\xi*\H-\H/2});

% Polar domain at each time
\draw[color=black!\colorfrac,domain=0:90, smooth, variable=\phi] plot ({\shiftright+(\H-(\H-\W)*\timefrac*(3/8+1/8*cos(2*\phi)))*cos(\phi)-\H/2}, {(\H-(\H-\W)*\timefrac*(3/8+1/8*cos(2*\phi)))*sin(\phi)-\H/2}) --plot ({\shiftright+\W*cos(90-\phi)-\H/2}, {\W*sin(90-\phi)-\H/2}) -- cycle;

% Polar arrows at each time
\draw[color=black!\colorfrac, draw opacity=0,domain=0:18, smooth, variable=\phi,->,>=stealth] plot ({\shiftright+(\H-(\H-\W)*\timefrac*(3/8+1/8*cos(2*\phi)))*cos(\phi)-\H/2}, {(\H-(\H-\W)*\timefrac*(3/8+1/8*cos(2*\phi)))*sin(\phi)-\H/2});
\draw[color=black!\colorfrac, draw opacity=0,domain=0:36, smooth, variable=\phi,->,>=stealth] plot ({\shiftright+(\H-(\H-\W)*\timefrac*(3/8+1/8*cos(2*\phi)))*cos(\phi)-\H/2}, {(\H-(\H-\W)*\timefrac*(3/8+1/8*cos(2*\phi)))*sin(\phi)-\H/2});
\draw[color=black!\colorfrac, draw opacity=0,domain=0:54, smooth, variable=\phi,->,>=stealth] plot ({\shiftright+(\H-(\H-\W)*\timefrac*(3/8+1/8*cos(2*\phi)))*cos(\phi)-\H/2}, {(\H-(\H-\W)*\timefrac*(3/8+1/8*cos(2*\phi)))*sin(\phi)-\H/2});
\draw[color=black!\colorfrac, draw opacity=0,domain=0:72, smooth, variable=\phi,->,>=stealth] plot ({\shiftright+(\H-(\H-\W)*\timefrac*(3/8+1/8*cos(2*\phi)))*cos(\phi)-\H/2}, {(\H-(\H-\W)*\timefrac*(3/8+1/8*cos(2*\phi)))*sin(\phi)-\H/2});

} % End time loop

% Cartesian Gammas
\node[color=black!40, anchor=north,rotate=90] at (\shiftleft+\W/2,    0) {$\phantom{_s}\Gamma_s\phantom{_0}$}; % Gamma_s
\node[anchor=south]                           at (\shiftleft     , \H/2) {$\Gamma_0$}; % Top
\node[anchor=north]                           at (\shiftleft     ,-\H/2) {$\Gamma_0$}; % Bottom
\node[anchor=south,rotate=90]                 at (\shiftleft-\W/2,    0) {$\phantom{_0}\Gamma_0\phantom{_0}$}; % Left

\node[color=black!100, anchor=south,rotate={atan2(\H,-\W/4*pi*cos(90*.5))}] at ({\shiftleft+\W/2 -\W/4*(1+2*sin(90*.5))},{.5*\H-\H/2})   {$\phantom{_s}\Gamma_s\phantom{_0}$}; % Gamma_s after

\node[color=black!40, anchor=west,rotate=0] at (\shiftleft+\W/2,-\H*.4)   {$\xi$}; % xi before
\node[color=black!100, anchor=east,rotate={atan2(\H,-\W/4*pi*cos(90*0))-90}] at ({\shiftleft+\W/2 -\W/4*(1+2*sin(90*.1))},{.1*\H-\H/2})   {$\xi$}; % xi after

% Polar

\node[color=black!40,anchor=south west]      at ( {\shiftright+ (\H)*sqrt(2)/2-\H/2},{(\H)*sqrt(2)/2-\H/2})   {$\Gamma_s$}; % Outer
\node[               anchor=north east]      at (  \shiftright+ -\H/8,-\H/8)  {$\Gamma_0$}; % Inner
\node[               anchor=north]           at (  \shiftright+  \H/4,-\H/2)  {$\Gamma_0$}; % Bottom
\node[               anchor=south,rotate=90] at (  \shiftright+ -\H/2, \H/4)  {$\Gamma_0$}; % Left

\node[color=black!100,anchor=north east]     at ({\shiftright+(\H-(\H-\W)*(3/8+1/8*cos(2*45)))*cos(45)-\H/2}, {(\H-(\H-\W)*(3/8+1/8*cos(2*45)))*sin(45)-\H/2})   {$\Gamma_s$}; % Outer

\node[color=black!100,anchor=east]     at ({\shiftright+(\H-(\H-\W)*(3/8+1/8*cos(2*5)))*cos(5)-\H/2}, {(\H-(\H-\W)*(3/8+1/8*cos(2*5)))*sin(5)-\H/2})   {$\xi$}; % xi after
\node[color=black!40, anchor=west]     at ({\shiftright+(\H-(\H-\W)*0*(3/8+1/8*cos(2*5)))*cos(5)-\H/2}, {(\H-(\H-\W)*0*(3/8+1/8*cos(2*5)))*sin(5)-\H/2})   {$\xi$}; % xi before
\end{tikzpicture}%}
\caption{\reviewerThree{The boundary $\Gamma$ of the domain consists of an ablating surface $\Gamma_s$ and a non-ablating surface $\Gamma_0$, such that $\Gamma=\Gamma_s \cup \Gamma_0$.  $\Gamma_s$ is parameterized by $\xi\in[0,1]$, which increases in the counterclockwise direction.}}
\label{fig:surfaces}
\end{figure}
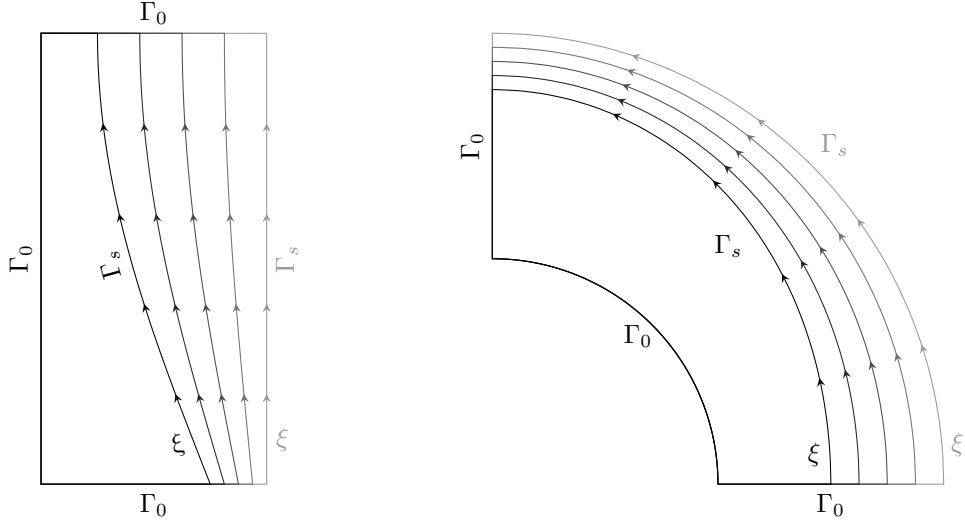

We denote the time-dependent domain of the material by $\Omega(t)$.  The boundary $\Gamma$ of the domain consists of an ablating surface $\Gamma_s$ and a non-ablating surface $\Gamma_0$, such that $\Gamma=\Gamma_s \cup \Gamma_0$. 
We denote the ablating surface by $\Gamma_s=\{(x,y): x=x_s,\,y=y_s\}$, which is arbitrarily parameterized by $\mathbf{x}_s(\xi,t)=(x_s(\xi,t),y_s(\xi,t))$, where $\xi\in[0,1]$ increases in the counterclockwise direction, and $t\in[0,\,\bar{t}]$, \reviewerTwo{with $\bar{t}$ denoting the final time}.  \reviewerThree{Figure~\ref{fig:surfaces} shows examples of $\Gamma$ and $\xi$.}

Along the ablating surface $\Gamma_s$, the material recedes by an amount $s(\xi,t)$ in the direction opposite to the outer normal of the surface, such that the recession rate is defined by
\begin{align}
\dot{s}(\xi,t)=-\frac{\partial\mathbf{x}_s}{\partial t}(\xi,t) \cdot \mathbf{n}_s(\xi,t),
\label{eq:s_dot_def}
\end{align}
where the outer unit normal vector is defined by
\begin{align}
\mathbf{n}_s(\xi,t) = \frac{1}{\sqrt{\left(\partial x_s/\partial \xi\right)^2+\left(\partial y_s/\partial \xi\right)^2}}\frac{\partial}{\partial \xi}\left\{\begin{matrix*}[r]y_s\\-x_s\end{matrix*}\right\}.
\label{eq:n_s}
\end{align}
%
%Appendix~\ref{sec:app_a} provides a brief discussion on the implications of the parameter choice $\xi$.

The recession rate is modeled by 
\begin{align}
\dot{s}(\xi,t)=\frac{B'(T_s,\reviewerTwo{p_e})\reviewerTwo{C_e}}{\reviewerTwo{\rho}},
\label{eq:s_dot}
\end{align}
where \reviewerTwo{%
$T_s(\xi,t)=T(\mathbf{x}_s(\xi,t),t)$ is the temperature of the solid along the ablating surface, 
$p_e(\xi,t)$ is the pressure at the outer edge of the boundary layer expressed in terms of position along the ablating surface, and
$B'(T_s,p_e)$ is the nondimensionalized char ablation rate. 
The heat transfer coefficient, $C_e(\xi,t)$, is commonly denoted by $\rho_e u_e C_h$, where $C_h$ is the Stanton number and $\rho_e$ and $u_e$ are the density and velocity at the outer edge of the boundary layer~\cite{amar_2008}}.

Defining $q_s=\mathbf{q}_s\cdot\mathbf{n}_s$, the heat flux \reviewerThree{normal to} the ablating surface is
\begin{align}
q_s = %%\nonumber\\
\reviewerTwo{C_e}\left[h_w(T_s,\reviewerTwo{p_e})-h_r\right] + \reviewerTwo{\rho \dot{s}\left[h_w(T_s,\reviewerTwo{p_e})-h_s(T_s)\right]} + \reviewerThree{\epsilon\sigma\left(T_s^4 - T_r^4\right)}.
\label{eq:heat_flux}
\end{align}
In~\eqref{eq:heat_flux}, the first term is the convective heat flux, the second term is the energy loss from ablation, \reviewerThree{and the third term is the radiative flux}.  $h_w(T_s,\reviewerTwo{p_e})$ is the wall enthalpy and $h_r(\xi,t)$ is the recovery enthalpy. $h_s(\xi,t)$ is the solid enthalpy, computed from
\begin{align}
h_s(T_s) = h_0+\int_{T_0}^{T_s} c_p(\hat{T})d\hat{T},
\label{eq:h_s}
\end{align}  
where, for our purposes, we set $h_0=0$~J/kg and $T_0=273.15$~K.  \reviewerThree{$\epsilon$ is the emissivity, $\sigma$ is the Stefan--Boltzmann constant, and $T_r$ is a radiation reference temperature, which we set to $T_r=300$~K.}
$B’(T_s,\reviewerTwo{p_e})$ in~\eqref{eq:s_dot} and $h_w(T_s,\reviewerTwo{p_e})$ in~\eqref{eq:heat_flux} are both computed from a surface thermochemistry model and provided as tabulated data, \reviewerThree{and $\epsilon$ in~\eqref{eq:heat_flux} is modeled as a constant}.  

From~\eqref{eq:fourier} and referencing~\eqref{eq:heat_flux}, the boundary condition along the ablating surface is
\begin{align}
-k(T_s)\frac{\partial T}{\partial n} =  q_s.
\label{eq:bc_ablating}
\end{align}
%
%
%\begin{align}
%\left.-k(T)\frac{\partial T}{\partial n}\right|_{s(t)} = \left.k(T)\frac{\partial T}{\partial z}\right|_{s(t)} = \tilde{q}(t),  
%\label{eq:left_q}
%\end{align}
%%
%where 
%%
%%\begin{align}
%.  To simplify the notation, in this paper, the vertical bar denotes the evaluation of each of the factors in the accompanying term at the spatial location specified in the subscript, throughout time.
%\label{eq:q_tilde}
%\end{align}
%
\reviewerThree{The non-ablating surface $\Gamma_0$ is insulated, such that there is no heat flux ($q_0=\mathbf{q}_0\cdot\mathbf{n}_0=0$), and}, from~\eqref{eq:fourier}, 
%\begin{align*}
%\mathbf{q}\cdot\mathbf{n} = -k(T)\nabla T \cdot \mathbf{n} = - k(T)\frac{\partial T}{\partial n} = 0,
%%\label{eq:right_q}
%\end{align*}
%%
%such that
%
\begin{align}
\frac{\partial T}{\partial n} = 0.
\label{eq:bc_nonablating}
\end{align}
%

%==============================================================================
\section{Manufactured Solutions} %=============================================
%==============================================================================
\label{sec:mms}

A governing system of partial differential equations can be written generally as
\begin{align}
\mathbf{r}(\mathbf{u};\boldsymbol{\mu})=\mathbf{0},
\label{eq:orig}
\end{align}
where \reviewerThree{$\mathbf{r}(\mathbf{u};\boldsymbol{\mu})$ is a symbolic operator representing the governing equations}, $\mathbf{u}=\mathbf{u}(\mathbf{x},t)$ is the state vector, and $\boldsymbol{\mu}$ is the parameter vector.
To solve~\eqref{eq:orig} numerically, it must be discretized in time and space:
\begin{align*}
\mathbf{r}_h(\mathbf{u}_h;\boldsymbol{\mu})=\mathbf{0},
%\label{eq:disc}
\end{align*}
where $\mathbf{r}_h$ is the residual of the discretized system of equations, and $\mathbf{u}_h$ is the solution to the discretized equations.

\reviewerThree{There is an a priori error estimate for the} discretization error $\mathbf{e}_\mathbf{u} = \mathbf{u}_h - \mathbf{u}$ and its norm, \reviewerThree{which asymptotically has the form}  $\|\mathbf{e}_\mathbf{u}\|\approx C h^p$, where $C$ is a function of the solution derivatives, $h$ is representative of the discretization size, and $p$ is the order of accuracy.  Through convergence studies of the norm of the error, we can assess whether the expected order of accuracy is obtained.

However, $\mathbf{e}_\mathbf{u}$ can only be measured if $\mathbf{u}$ is known.  Exact solutions to~\eqref{eq:orig} require negligible implementation effort, but are generally too limited to fully exercise the capabilities of the code.  Manufactured solutions are therefore popular alternatives, which typically introduce a forcing vector into the original equations to coerce the solution to the manufactured one:
\begin{align}
\mathbf{r}_h(\mathbf{u}_h;\boldsymbol{\mu})=\mathbf{r}(\mathbf{u}_\text{MS};\boldsymbol{\mu}).
\label{eq:mms}
\end{align}
In~\eqref{eq:mms}, $\mathbf{r}(\mathbf{u}_\text{MS};\boldsymbol{\mu})$ is computed analytically since $\mathbf{r}$, $\mathbf{u}_\text{MS}$, and $\boldsymbol{\mu}$ are known.  \reviewerThree{To simplify the notation, we have assumed the appropriate mapping of $\mathbf{r}(\mathbf{u}_\text{MS};\boldsymbol{\mu})$ onto the discrete space, where it is evaluated.}

An alternative approach, which we employ in this paper, involves manufacturing the parameters instead to obtain $\mathbf{r}(\mathbf{u};\boldsymbol{\mu}_{\text{MP}})=\mathbf{0}$, which is solved numerically by
\begin{align}
\mathbf{r}_h(\mathbf{u}_h;\boldsymbol{\mu}_{\text{MP}})=\mathbf{0}.
\label{eq:mmp}
\end{align}
Unlike the approach in~\eqref{eq:mms}, the approach in~\eqref{eq:mmp} does not require code modification. 

To compute $\mathbf{u}$, we derive solutions to the governing equations.  For the boundary conditions that cannot be satisfied, we manufacture the underlying parameters $\boldsymbol{\mu}_{\text{MP}}$.  For ablation, we demonstrate this approach in Sections~\ref{sec:solutions} and~\ref{sec:bc}.

\section{Heat Equation Solution}
\label{sec:solutions}

We consider the temperature-dependent material properties $k(T) = \bar{k} f(T)$ and $c_p(T) = \bar{c}_pf(T)$, where $f(T)>0$ for $T>0$.  Employing a Kirchhoff transformation,

\begin{align}
\theta(\mathbf{x},t) &{}= \frac{1}{\bar{k}}\int_{T} k(T') dT' + C_k = \int_{T} f(T') dT' + C_k  = F(T),
\label{eq:T_to_theta}
\end{align}
where $\theta$ denotes the transformed temperature, we obtain
\begin{align}
\frac{\partial \theta}{\partial t} &{}= f(T) \frac{\partial T}{\partial t}, \nonumber \\[0.5em]
\nabla \theta &{}= f(T) \nabla T \label{eq:grad_theta},
\end{align}
which are substituted into~\eqref{eq:energy_2} to yield a constant-coefficient heat equation
\begin{align}
\frac{\partial \theta}{\partial t}   - \bar{\alpha}\Delta\theta = 0.
\label{eq:theta_energy}
\end{align}
From~\eqref{eq:grad_theta}, the normal derivatives are related by
\begin{align}
\frac{\partial T}{\partial n}=\frac{1}{f(T)}\frac{\partial\theta}{\partial n}. \label{eq:dtheta_dn}
\end{align}

\reviewerTwoThree{%
While a constant thermal diffusivity significantly simplifies the solution to the heat equation, limited analytic solutions can be obtained from other functional forms for $k$ and $c_p$.  Nonetheless, because $f(T)$ permits $k$ and $c_p$ to vary with $T$ and test that capability of the code, we restrict $k$ and $c_p$ to these forms.
%The functional forms of $k$ and $c_p$ are necessary to obtain
}

To solve~\eqref{eq:theta_energy}, we temporarily disregard the time dependency of the domain \reviewerThree{due to the ablating surface} and assume we can separate the time and space dependencies of the solution, such that
%
%\begin{align}
%%T_h(\mathbf{x},t)=\sum_{j=0}^\infty\hat{T}_j(t)\varphi_j(\mathbf{x}).
%\theta(\mathbf{x},t)=\hat{\theta}(t)\varphi(\mathbf{x}).
%\label{eq:T_h_single}
%\end{align}
\begin{align}
\theta(\mathbf{x},t) = \reviewerThree{\sum_{i=0}^\infty\sum_{j=0}^\infty}\hat{\theta}_{i,j}(t) \varphi_{i,j}(\mathbf{x}),
\label{eq:full_theta_form}
\end{align}
where $\varphi_{i,j}(\mathbf{x})$ is an orthogonal basis, and $i$ and $j$ are indices associated with the basis of different spatial coordinates.
Substituting~\eqref{eq:full_theta_form} into~\eqref{eq:theta_energy} yields
\begin{align}
\frac{1}{\bar{\alpha}}\frac{\hat{\theta}_{i,j}'(t)}{\hat{\theta}_{i,j}(t)} = \frac{\Delta \varphi_{i,j}(\mathbf{x})}{\varphi_{i,j}(\mathbf{x})} = -\lambda_{i,j}.
\label{eq:heat_3h}
\end{align}

From~\eqref{eq:heat_3h}, 
\begin{align}
\hat{\theta}_{i,j}(t) = \hat{\theta}_{{i,j}_0} e^{-\bar{\alpha}\lambda_{i,j}t},
\label{eq:theta_time}
\end{align}
where
\begin{align*}
\hat{\theta}_{{i,j}_0} = \frac{\displaystyle\int_{\Omega(0)} \theta(\mathbf{x},0) \varphi_{i,j}(\mathbf{x}) d\Omega}{\displaystyle\int_{\Omega(0)} \varphi_{i,j}(\mathbf{x})^2 d\Omega}.
\end{align*}
Because we are focusing on ablative processes and interested in verifying the time integrator, we are particularly interested in cases where the temperature increases with time, which occurs when $\lambda_{i,j}<0$.

In the following two subsections, we derive $\varphi_{i,j}(\mathbf{x})$ and $\lambda_{i,j}$ for 1) a particular type of domain in Cartesian coordinates, and 2) a particular type of domain in polar coordinates.

\subsection{Cartesian Coordinates}
\label{sec:cartesian}

\begin{figure}%[!b]
\centering
\begin{tikzpicture}
%\useasboundingbox (-7,-4) (7,4);
\path (-4, 0) -- ( 4, 0);
\path ( 0,-4.1) -- ( 0, 4.1);
\def\H{6}
\def\W{3}
\def\tbar{5}
\def\nscale{.7}
%\draw (-\W/2,-\H/2) -- (\W/2,-\H/2) -- (\W/2,\H/2) -- (-\W/2,\H/2) -- (-\W/2,-\H/2);
\def\nincr{4}
\foreach \i in {0,...,\nincr}%
{
\def\timefrac{\i/\nincr}
\def\maxblack{40}
\def\minblack{100}
\pgfmathsetmacro{\colorfrac}{\i/\nincr*\minblack + (\nincr-\i)/\nincr*\maxblack}
%\colorlet{graycolor}{gray}{(\i+1)/(\nincr+1)}

% Domain at each time
\draw[color=black!\colorfrac,domain=0:1, smooth, variable=\xi] plot ({\W/2 -\W/4*\timefrac*(1+2*sin(90*\xi))},{\xi*\H-\H/2}) -- (-\W/2,\H/2) -- (-\W/2,-\H/2) -- cycle;

% Normal vector
\draw[color=black!\colorfrac,->,>=stealth, ] ({-\W/2+(1-1/4*\timefrac*(1+sqrt(2)))*\W},0) -- ({-\W/2+(1-1/4*\timefrac*(1+sqrt(2)))*\W + \nscale*\H/sqrt(\H^2+pi^2*\timefrac^2*\W^2/32)},{\nscale*pi*\timefrac*\W/sqrt(32*\H^2+pi^2*\timefrac^2*\W^2)}) ;
}
%\draw[gray,domain=0:1, smooth, variable=\xi] plot ({\W/2-  \W/8*sin(45*(\xi+1))^2},{\xi*\H-\H/2});
%\draw[gray,domain=0:1, smooth, variable=\xi] plot ({\W/2-  \W/4*sin(45*(\xi+1))^2},{\xi*\H-\H/2});
%\draw[gray,domain=0:1, smooth, variable=\xi] plot ({\W/2-3*\W/8*sin(45*(\xi+1))^2},{\xi*\H-\H/2});
%\draw[domain=0:1, smooth, variable=\xi]      plot ({\W/2-  \W/2*sin(45*(\xi+1))^2},{\xi*\H-\H/2});
%\node[anchor=north,rotate=90] at (\W/2+\nscale,0)   {$\displaystyle\frac{\partial T}{\partial n}(\mathbf{x}_s,t)=-\frac{1}{k(T_s)}\mathbf{q}_s\cdot\mathbf{n}_s$};%a(\xi,t) g(T(\mathbf{x}_s,t)) + b(\xi,t)$}; % Right
\node[anchor=north,rotate=90] at (\W/2+\nscale,0)   {$\displaystyle -k(T_s)\frac{\partial T}{\partial n}(\mathbf{x}_s,t)=q_s$}; % Right
\node[anchor=south] at (0,\H/2)  {$\displaystyle\frac{\partial T}{\partial y}(x,H,t)=0$}; % Top
\node[anchor=north] at (0,-\H/2) {$\displaystyle\frac{\partial T}{\partial y}(x,0,t)=0$}; % Bottom
\node[anchor=south,rotate=90] at (-\W/2,0)  {$\displaystyle\frac{\partial T}{\partial x}(0,y,t)=0$}; % Left
%\draw[->,>=stealth, ] (0.0764145*\W,-0.0113909*\H) -- ({0.0764145*\W+\H/sqrt(\H^2+.282604^2*\W^2)},{-0.0113909*\H+.282*\W/sqrt(\H^2+.282604^2*\W^2)});
%\draw[->,>=stealth, ] (\W/2,0) -- (\W/2+1,0);
\end{tikzpicture}%}
\caption{Domain and boundary conditions in Cartesian coordinates.}
\label{fig:cartesian_domain}
\end{figure}
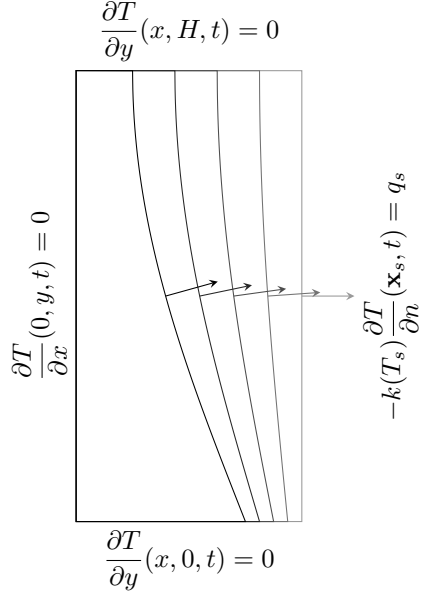

For this problem, the domain is defined by $\Omega = \{(x,y): 0\le x\le \reviewerThree{x_s(y)},\,0\le  y \le H\}$.  The ablating surface $\Gamma_s$ is subject to the boundary condition in~\eqref{eq:bc_ablating}, whereas the remaining edges ($x=0$, $y=0$, $y=H$) comprise $\Gamma_0$, with the boundary condition in~\eqref{eq:bc_nonablating}.  Figure~\ref{fig:cartesian_domain} provides an example of this domain; however, it is not necessary for the domain to initially be a rectangle.  The requirement is that the three edges that comprise $\Gamma_0$ remain straight, \reviewerTwo{with the middle edge perpendicular to the mutually parallel remaining edges.} %and mutually perpendicular.

With this requirement, we assume we can separate the $x$ and $y$ dependencies, such that 
\begin{align}
\varphi_{i,j}(\mathbf{x})=u_i(x) v_j(y).
\label{eq:cartesian_phi}
\end{align}
From~\eqref{eq:heat_3h}, we obtain
\begin{align}
\lambda_{i,j} +\frac{u_i''(x)}{u_i(x)} =  -\frac{v_j''(y)}{v_j(y)} = \nu_j^2.
\label{eq:heat_4h}
\end{align}

From~\eqref{eq:heat_4h}, $v_j''(y) + \nu_j^2 v_j(y) = 0$, and from~\eqref{eq:bc_nonablating} and~\eqref{eq:dtheta_dn}, at $y=0$ and $y=H$, 
$\frac{\partial\theta}{\partial y}=0$, such that $v_j'(0)=v_j'(H)=0$.  For a nontrivial solution for $v_j$, $\nu_j =j\pi/H$ for $j\in\mathbb{N}_0$, such that
\begin{align}
v_j(y) = \cos(j\pi y/H). %C_{v_j}
\label{eq:v_j}
\end{align}

From~\eqref{eq:heat_4h}, $u_i''(x) + \mu_i^2 u_i(x) = 0$, where $\mu_i^2=\lambda_{i,j}-\nu_j^2$, and $\mu_i^2$ is real.  From~\eqref{eq:bc_nonablating} and~\eqref{eq:dtheta_dn}, at $x=0$, $\frac{\partial \theta}{\partial x}=0$, such that $u_i'(0)=0$.  For a nontrivial solution for $u_i$,
\begin{align}
u_i(x) = \left\{
\begin{matrix*}[r]
\cosh(|\mu_i| x) & \text{for } \mu_i^2<0
\\
\cos (\phantom{|}\mu_i\phantom{|} x) & \text{for } \mu_i^2\ge0
\end{matrix*}
\right., %C_{u_i}
\label{eq:u_i}
\end{align}
where $\mu_i$ depends on the boundary condition at $x=x_s$.  %If heat were not entering the system, $u_0(x)$ would instead be zero.

%As a result,
%%
%\begin{align}
%\varphi_{i,j}(\mathbf{x}) = u_i(x) v_j(y),
%\label{eq:phi}
%\end{align}
%%
%such that
%
%\begin{align}
%\theta(\mathbf{x},t) = \sum_{i,j=0}^\infty\hat{\theta}_{i,j}(t) \varphi_{i,j}(\mathbf{x}),
%\label{eq:full_theta_form}
%\end{align}
%

%
%where
%%
%\begin{align}
%\lambda_{i,j} &{}= \left\{
%\begin{matrix}
%-\mu_i^2 + \nu_j^2 & \text{for } i=0
%\\[0.5em]
%\phantom{-}\mu_i^2 + \nu_j^2 & \text{for } i>0
%\end{matrix}
%\right.,
%\label{eq:lambda}
%\end{align}
%%
%and
%%

With $\varphi_{i,j}(\mathbf{x})$~\eqref{eq:cartesian_phi} and 
\begin{align}
\lambda_{i,j} = \mu_i^2 + \nu_j^2
\label{eq:cartesian_lambda}
\end{align}
known, \eqref{eq:full_theta_form} satisfies~\eqref{eq:theta_energy} and the boundary conditions on $\Gamma_0$~\eqref{eq:bc_nonablating}.  However, the boundary condition on $\Gamma_s$~\eqref{eq:bc_ablating} has not been addressed and $\mu_i$ has not been determined.
Because the domain varies with respect to time and $x_s$ can vary with respect to $y$, we cannot satisfy general boundary conditions with~\eqref{eq:full_theta_form}.  Therefore, in Section~\ref{sec:bc}, we manufacture the ablating boundary condition such that it is always satisfied.

\subsection{Polar Coordinates}

\begin{figure}%[!b]
\centering
%\frame{%
\begin{tikzpicture}
%\useasboundingbox (-7,-4) (7,4);
\path (-7, 0) -- ( 7, 0);
\path ( 0,-4.1) -- ( 0, 4.1);
\def\H{6}
\def\W{3}
\def\tbar{5}
\def\nscale{.7}
%\draw (-\W/2,-\H/2) -- (\W/2,-\H/2) -- (\W/2,\H/2) -- (-\W/2,\H/2) -- (-\W/2,-\H/2);
\def\nincr{4}
\foreach \i in {0,...,\nincr}%
{
\def\timefrac{\i/\nincr}
\def\maxblack{40}
\def\minblack{100}
\pgfmathsetmacro{\colorfrac}{\i/\nincr*\minblack + (\nincr-\i)/\nincr*\maxblack}

\draw[color=black!\colorfrac,domain=0:90, smooth, variable=\phi] plot ({(\H-(\H-\W)*\timefrac*(3/8+1/8*cos(2*\phi)))*cos(\phi)-\H/2}, {(\H-(\H-\W)*\timefrac*(3/8+1/8*cos(2*\phi)))*sin(\phi)-\H/2}) --plot ({\W*cos(90-\phi)-\H/2}, {\W*sin(90-\phi)-\H/2}) -- cycle;

% Normal vector
\draw[color=black!\colorfrac,->,>=stealth, ] ({-\H/2+(\H-(3*\timefrac*(\H-\W))/8)/sqrt(2)},{-\H/2+(\H-(3*\timefrac*(\H-\W))/8)/sqrt(2)}) -- ({-\H/2+(\H-(3*\timefrac*(\H-\W))/8)/sqrt(2)+(\nscale*(-(\H*(-8+\timefrac))+\timefrac*\W))/sqrt((\H*(-8+5*\timefrac)-5*\timefrac*\W)^2+(\H*(-8+\timefrac)-\timefrac*\W)^2)},{-\H/2+(\H-(3*\timefrac*(\H-\W))/8)/sqrt(2)-(\nscale*(\H*(-8+5*\timefrac)-5*\timefrac*\W))/sqrt((\H*(-8+5*\timefrac)-5*\timefrac*\W)^2+(\H*(-8+\timefrac)-\timefrac*\W)^2)}) ;
}
%
%\node[anchor=south] at ({(\H+1.5*\nscale)*sqrt(2)/2-\H/2},{(\H+1.5*\nscale)*sqrt(2)/2-\H/2})   {$\displaystyle\frac{\partial T}{\partial n}(\mathbf{x}_s,t)=-\frac{1}{k(T_s)}\mathbf{q}_s\cdot\mathbf{n}_s$}; % Outer
\node[anchor=south] at ({(\H+1.25*\nscale)*sqrt(2)/2-\H/2},{(\H+1.25*\nscale)*sqrt(2)/2-\H/2})   {$\displaystyle -k(T_s)\frac{\partial T}{\partial n}(\mathbf{x}_s,t)=q_s$}; % Outer
\node[anchor=north east] at (-\H/8,-\H/8)  {$\displaystyle\frac{\partial T}{\partial r}(r_0,\phi,t)=0$}; % Inner
\node[anchor=north] at (\H/4,-\H/2) {$\displaystyle\frac{\partial T}{\partial \phi}(r,0,t)=0$}; % Bottom
\node[anchor=south,rotate=90] at (-\H/2,\H/4)  {$\displaystyle\frac{\partial T}{\partial \phi}(r,\bar{\phi},t)=0$}; % Left

\end{tikzpicture}%}
\caption{Domain and boundary conditions in polar coordinates.}
\label{fig:polar_domain}
\end{figure}
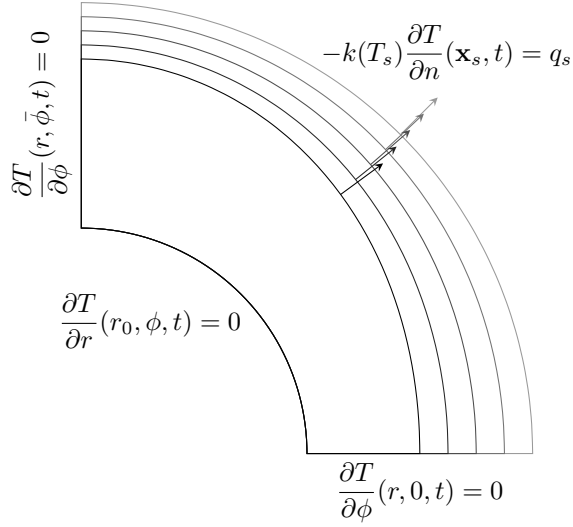

For this problem, the domain is defined by $\Omega = \{(r,\phi): r_0\le r\le \reviewerThree{r_s(\phi)},\,0\le  \phi \le \bar{\phi}\}$.  The ablating surface $\Gamma_s$ is subject to the boundary condition in~\eqref{eq:bc_ablating}, whereas the remaining edges ($r=r_0$, $\phi=0$, $\phi=\bar{\phi}$) comprise $\Gamma_0$, with the boundary condition in~\eqref{eq:bc_nonablating}.  Figure~\ref{fig:polar_domain} provides an example of this domain with $\bar{\phi}=\pi/2$; however, it is not necessary for the domain to initially be a fractional annulus.  The requirements are that the two radial edges that comprise $\Gamma_0$ remain radial at fixed angles, and the inner \reviewerTwo{edge remains a circular arc}.

With these requirements, we assume we can \reviewerTwo{partially} separate the $r$ and $\phi$ dependencies, such that 
\begin{align*}
\varphi_{i,j}(\mathbf{x})=u_{i,j}(r) v_j(\phi).
%\label{eq:polar_phi}
\end{align*}
Note that, unlike $u_i(x)$ and $v_j(y)$, which are decoupled in Section~\ref{sec:cartesian}, $u_{i,j}(r)$ depends on $v_j(\phi)$.  From~\eqref{eq:heat_3h}, we obtain
\begin{align}
\lambda_{i,j} r^2 +\frac{r^2 u_{i,j}''(r)+r u_{i,j}'(r)}{u_{i,j}(r)} =  -\frac{v_j''(\phi)}{v_j(\phi)} = \nu_j^2.
\label{eq:heat_4h_polar}
\end{align}

From~\eqref{eq:heat_4h_polar}, $v_j''(\phi) + \nu_j^2 v_j(\phi) = 0$, and from~\eqref{eq:bc_nonablating} and~\eqref{eq:dtheta_dn}, at $\phi=0$ and $\phi=\bar{\phi}$, 
$\frac{\partial\theta}{\partial \phi}=0$, such that $v_j'(0)=v_j'(\bar{\phi})=0$.  For a nontrivial solution for $v_j$, $\nu_j =j\pi/\bar{\phi}$ for $j\in\mathbb{N}_0$, such that
\begin{align}
v_j(\phi) = \cos(j\pi \phi/\bar{\phi}). %C_{v_j}
\label{eq:v_j_polar}
\end{align}

From~\eqref{eq:heat_4h_polar}, we obtain
\begin{align*}
r^2 u_{i,j}''(r) + r u_{i,j}'(r) + (\lambda_{i,j} r^2 - \nu_j^2) u_{i,j}(r) = 0,
%\label{eq:modified_bessel}
\end{align*}
%
%For ablation problems, the temperature is expected to increase with time, such that $\lambda$~\eqref{eq:theta_time} is negative.  Therefore, \eqref{eq:modified_bessel} is a modified Bessel equation and its solutions are comprised of modified Bessel functions of the first and second kind.
%
and from~\eqref{eq:bc_nonablating} and~\eqref{eq:dtheta_dn}, at $r=r_0$, $\frac{\partial \theta}{\partial r}=0$, such that $u_{i,j}'(r_0)=0$.  For a nontrivial solution for $u_{i,j}$,
%
%\begin{widetext}
%\begin{align}
%u_{i,j}(r) =  \left\{
%\begin{matrix*}[r]
%K_{i,j} I_{\nu_j}(\sqrt{|\lambda_{i,j}|}r) + I_{i,j} K_{\nu_j}(\sqrt{|\lambda_{i,j}|}r) & \text{for } \lambda_{i,j}<0 \\[.5em]
%Y_{i,j} J_{\nu_j}(\sqrt{\phantom{|}\lambda_{i,j}\phantom{|}}r) + J_{i,j} Y_{\nu_j}(\sqrt{\phantom{|}\lambda_{i,j}\phantom{|}}r) & \text{for } \lambda_{i,j}>0 \\[.5em]
%\cosh(\nu_j \ln (r/r_0))  & \text{for } \lambda_{i,j}=0
%\end{matrix*}
%\right., %C_{u_i}
%\label{eq:u_in}
%\end{align}
%\end{widetext}
%
\begin{align}
u_{i,j}(r) = %\nonumber \\  
\left\{
\begin{matrix*}[r]
K_{i,j} I_{\nu_j}(r') + I_{i,j} K_{\nu_j}(r') & \text{for } \lambda_{i,j}<0 \\[.5em]
Y_{i,j} J_{\nu_j}(r') + J_{i,j} Y_{\nu_j}(r') & \text{for } \lambda_{i,j}>0 \\[.5em]
\cosh(\nu_j \ln (r/r_0))  & \text{for } \lambda_{i,j}=0
\end{matrix*}
\right., %C_{u_i}
\label{eq:u_in}
\end{align}
where $r'=\sqrt{|\lambda_{i,j}|}r$, $I_\alpha$ and $K_\alpha$ are modified Bessel functions of the first and second kind~\cite{abramowitz_1964}, $J_\alpha$ and $Y_\alpha$ are Bessel functions of the first and second kind, and
%
%\begin{alignat*}{7}
%K_{i,j} &{}={}&  K_{\nu_j-1}\left(\sqrt{|\lambda_{i,j}|}r_0\right) &{}+{}& K_{\nu_j+1}\left(\sqrt{|\lambda_{i,j}|}r_0\right), \\
%I_{i,j} &{}={}&  I_{\nu_j-1}\left(\sqrt{|\lambda_{i,j}|}r_0\right) &{}+{}& I_{\nu_j+1}\left(\sqrt{|\lambda_{i,j}|}r_0\right), \\
%Y_{i,j} &{}={}&  Y_{\nu_j-1}\left(\sqrt{\phantom{|}\lambda_{i,j}\phantom{|}}r_0\right) &{}-{}& Y_{\nu_j+1}\left(\sqrt{\phantom{|}\lambda_{i,j}\phantom{|}}r_0\right), \\
%J_{i,j} &{}={}& -J_{\nu_j-1}\left(\sqrt{\phantom{|}\lambda_{i,j}\phantom{|}}r_0\right) &{}+{}& J_{\nu_j+1}\left(\sqrt{\phantom{|}\lambda_{i,j}\phantom{|}}r_0\right).
%\end{alignat*}
%
\begin{alignat*}{7}
K_{i,j} &{}={}&  K_{\nu_j-1}(r_0') &{}+{}& K_{\nu_j+1}(r_0'), \\
I_{i,j} &{}={}&  I_{\nu_j-1}(r_0') &{}+{}& I_{\nu_j+1}(r_0'), \\
Y_{i,j} &{}={}&  Y_{\nu_j-1}(r_0') &{}-{}& Y_{\nu_j+1}(r_0'), \\
J_{i,j} &{}={}& -J_{\nu_j-1}(r_0') &{}+{}& J_{\nu_j+1}(r_0').
\end{alignat*}
$\lambda_{i,j}$ depends on the boundary condition at $r=r_s$.

%\vspace{10em}

As with Section~\ref{sec:cartesian}, the boundary condition on $\Gamma_s$~\eqref{eq:bc_ablating} has not yet been addressed and $\lambda_{i,j}$ has not been determined.  In Section~\ref{sec:bc}, we manufacture the ablating boundary condition such that it is always satisfied.

%\subsubsection{Boundary Condition Mapping}
%
%We map the ablating surface boundary condition in~\eqref{eq:variable_ablating_bc} to the actual ablating surface boundary condition~\eqref{eq:bc_ablating} to obtain the relationship
%%
%\begin{align*} 
%a(\xi,t) g(T_s) + b(\xi,t)
%=-\frac{1}{k(T_s)} \left(\reviewerTwo{C_e}\left[h_w(T_s,\reviewerTwo{p_e})-h_r\right] + \reviewerTwo{\rho} \dot{s}h_w(T_s,\reviewerTwo{p_e})\right).
%\end{align*}

\section{Boundary Condition Reconciliation}
\label{sec:bc}

In this section, we manufacture the ablating boundary condition so that we can manufacture solutions without adding a source term.  In doing so, we have much freedom, provided the functions are sufficiently smooth.  Like traditional manufactured solutions, certain desirable properties of the underlying functions, such as a sufficient number of finite nontrivial derivatives and elementary function composition, take precedence over being physically realizable.

We begin by manufacturing $T(\mathbf{x},t)$, which requires manufacturing the material properties $k(T)$, $c_p(T)$, and $\rho$, as well as $\theta(\mathbf{x},t)$~\eqref{eq:full_theta_form}.  

To manufacture the material properties, we must manufacture $\bar{k}$, $\bar{c}_p$, \reviewerTwo{$\rho$}, and \reviewerThree{$\epsilon$}, as well as $f(T)$.  From $\bar{k}$, $\bar{c}_p$, and $\reviewerTwo{\rho}$, we can compute $\bar{\alpha}$ in~\eqref{eq:theta_time}.  $f(T)$ relates $\theta(\mathbf{x},t)$ and $T(\mathbf{x},t)$ and provides the dependencies of $k(T)$ and $c_p(T)$ on $T$; it should be manufactured such that the inverse of~\eqref{eq:T_to_theta}, $T(\mathbf{x},t) = F^{-1}(\theta)$, can be easily computed.  

To manufacture $\theta(\mathbf{x},t)$, we truncate the series in~\eqref{eq:full_theta_form} and we specify $\hat{\theta}_{{i,j}_0}$ in $\hat{\theta}_{i,j}(t)$~\eqref{eq:theta_time}.  If we are using Cartesian coordinates, we specify $\mu_i$, which appears in $u_i(x)$~\eqref{eq:u_i} and $\lambda_{i,j}$~\eqref{eq:cartesian_lambda}.  If we are using polar coordinates, we specify $\lambda_{i,j}$.  With $\theta(\mathbf{x},t)$ specified, we can compute the temperature from $T(\mathbf{x},t)=F^{-1}(\theta)$.

Next, we manufacture $\mathbf{x}_s(\xi,t)$, such that we can compute $\mathbf{n}_s(\xi,t)$ from~\eqref{eq:n_s}, then $\dot{s}(\xi,t)$ from~\eqref{eq:s_dot_def}.

With $T(\mathbf{x},t)$ and $\mathbf{x}_s(\xi,t)$ manufactured, the next step is to manufacture the parameters to satisfy the boundary condition on $\Gamma_s$~\eqref{eq:bc_ablating}:
\begin{align}
q_s = -k(T_s)\frac{\partial T}{\partial n} = \reviewerTwo{C_e}\left[h_w(T_s,\reviewerTwo{p_e})-h_r\right] + \reviewerTwo{\rho} \dot{s}\left[h_w(T_s,\reviewerTwo{p_e})-h_s(T_s)\right] + \reviewerThree{\epsilon\sigma\big(T_s^4 - T_r^4\big)},
\label{eq:variable_ablating_bc}
\end{align}
as well as the recession rate~\eqref{eq:s_dot}.
In~\eqref{eq:variable_ablating_bc}, $k(T_s)$, $T_s$, $\frac{\partial T}{\partial n}$, $\reviewerTwo{\rho}$, and $\dot{s}(\xi,t)$ have already been determined, and $h_s(\xi,t)=h_s(T_s(\xi,t))$ can be computed from~\eqref{eq:h_s} using $T(\mathbf{x},t)$, $\mathbf{x}_s(\xi,t)$, and $c_p(T)$.  Therefore, $\reviewerTwo{C_e}(\xi,t)$, $\reviewerTwo{p_e}(\xi,t)$, $h_w(T_s,\reviewerTwo{p_e})$, and $h_r(\xi,t)$ need to be determined.  In~\eqref{eq:s_dot}, $\dot{s}(\xi,t)$ has already been determined, such that $B'(T_s,\reviewerTwo{p_e})$ and $\reviewerTwo{C_e}(\xi,t)$ need to be determined.  Therefore, we next manufacture $B'(T_s,\reviewerTwo{p_e})$ and $\reviewerTwo{p_e}(\xi,t)$.

Using~\eqref{eq:s_dot}, \eqref{eq:variable_ablating_bc} can be written as
\begin{align}
q_s = \reviewerTwo{C_e}\big(
h_w(T_s,\reviewerTwo{p_e})\big[1+ B'(T_s,\reviewerTwo{p_e})\big]     -h_r - B'(T_s,\reviewerTwo{p_e})h_s(T_s)
\big) + \reviewerThree{\epsilon\sigma\big(T_s^4 - T_r^4\big)}.
\label{eq:bc_ablating2}
\end{align}
In manufacturing the parameters, care must be taken to ensure that~\eqref{eq:bc_ablating2} does not introduce instabilities due to perturbations in the temperature, such as those that arise from discretization errors.  Therefore, as explained in~\ref{sec:app_b}, we impose $\frac{\partial q_s}{\partial T_s}\ge 0$.   
%
%For convenience, we consider the equality:
%In manufacturing the parameters, care must be taken to ensure that~\eqref{eq:bc_ablating2} does not change substantially from perturbations in the temperature, such as those that arise from discretization errors.  Therefore, to avoid such an instability, we cancel the dependence of $\mathbf{q}_s\cdot\mathbf{n}_s$ on $T_s$ by setting $\frac{\partial}{\partial T_s}(\mathbf{q}_s\cdot\mathbf{n}_s)=0$, such that
%
\reviewerThree{In~\eqref{eq:bc_ablating2},
\begin{align*}
\frac{\partial }{\partial T_s}\left[\epsilon\sigma\big(T_s^4 - T_r^4\big)\right] = 4\epsilon\sigma T_s^3 \ge 0.
\end{align*}
Therefore, $\frac{\partial q_s}{\partial T_s}\ge 0$ will be satisfied if we set
\begin{align*}
\frac{\partial }{\partial T_s}\left[\reviewerTwo{C_e}\big(
h_w(T_s,\reviewerTwo{p_e})\big[1+ B'(T_s,\reviewerTwo{p_e})\big]     -h_r - B'(T_s,\reviewerTwo{p_e})h_s(T_s)
\big)\right] =0,
\end{align*}
which yields
}
\begin{align*}
h_w(T_s,\reviewerTwo{p_e})\big[1+ B'(T_s,\reviewerTwo{p_e})\big]   - B'(T_s,\reviewerTwo{p_e})h_s(T_s) = g(\reviewerTwo{p_e}),
\end{align*}
\reviewerThree{such that}
\begin{align*}
h_w(T_s,\reviewerTwo{p_e}) = \frac{B'(T_s,\reviewerTwo{p_e})h_s(T_s)+g(\reviewerTwo{p_e})}{1+ B'(T_s,\reviewerTwo{p_e})}.
\end{align*}
For convenience, we set $g(\reviewerTwo{p_e})=0$.
With these functions known, we compute $\reviewerTwo{C_e}(\xi,t)$ from~\eqref{eq:s_dot} and $h_r(\xi,t)$ from~\eqref{eq:variable_ablating_bc}. 

\section{Numerical Examples}
\label{sec:results}

In this section, we demonstrate the methodology of Section~\ref{sec:bc} on problems in Cartesian and polar coordinate systems using \texttt{SIERRA Multimechanics Module:~Aria}~\cite{aria} for multiple discretizations.  The spatial domain is discretized using second-order-accurate finite elements, and the equations are integrated in time using a first-order-accurate backward Euler scheme.  Therefore, each subsequent discretization uses twice the number of elements in each \reviewerThree{spatial} dimension and \reviewerOneThree{a quarter of the time step size} as the previous discretization.  
Additionally, the piecewise linear interpolation of tabulated parameters is second-order accurate, such that, for each discretization, we \reviewerThree{halve the spacing between the data samples}.  \reviewerThree{Letting $h$ denote a quantity inversely proportional to the number of elements in one dimension and proportional to the square root of the time step size} and accounting for the aforementioned refinement ratios, we expect the error to be $\mathcal{O}(h^2)$.

We measure the error in the temperature using the norm
\begin{align}
\varepsilon_T = \max_{t\in[0,\,\bar{t}]} \big\|T_h(\mathbf{x},t)-T(\mathbf{x},t)\big\|_2,
\label{eq:error_norm_T}
\end{align}
by taking maximum over the time steps of the $L^2$-norm of the error over the spatial domain.  The subscript $h$ denotes the solution to the discretized equations.
We similarly measure the error in the ablating surface using the norm
\begin{align}
\varepsilon_{\mathbf{x}_s} = \max_{t\in[0,\,\bar{t}]} \big\|\mathbf{x}_{s_h}(\xi,t)-\mathbf{x}_s(\xi,t)\big\|_2.
\label{eq:error_norm_xs}
\end{align}
In~\eqref{eq:error_norm_xs}, the $L^2$-norm of the error is computed over the ablating surface.

Mesh deformation is accomplished through a Gent hyperelastic mesh stress model~\cite{gent_1996}.

\reviewerThree{For both problem sets, we consider cases with ($\epsilon=0.9$) and without ($\epsilon=0$) the radiative flux.}
For the material properties, we consider $\bar{\alpha}=\{10^{-8},\allowbreak 10^{-7},\allowbreak 10^{-6},\allowbreak 10^{-5}\}$~m$^2$/s, $\reviewerTwo{\rho}=1000$~kg/m$^3$, and $\bar{k} = 0.7$~W/m/K, which enable us to compute $\bar{c}_p$.  The multiple $\bar{\alpha}$ values enable us to consider different relative weights between the spatial and temporal contributions to the discretization error as we assess its convergence rate. For the thermochemical data, we manufacture 
\begin{align*}
B'(T_s,\reviewerTwo{p_e})=\exp\left(\frac{1}{1000}\frac{T_s}{\bar{T}} - \frac{1}{50}\frac{\reviewerTwo{p_e}}{\bar{p}}\right),
\end{align*}
where $\bar{T} = 1$~K, $\reviewerTwo{p_e}(t)=\bar{p} e^{5t/\bar{t}}/200$, $\bar{p}={}$101,325~Pa, and $\bar{t} = 5$~s.

%-------------------------------------------------------------------------------
\subsection{Cartesian Coordinates} %--------------------------------------------
%-------------------------------------------------------------------------------

The first problem set we consider uses Cartesian coordinates.  For the temperature dependence, we choose
\begin{align*}
f(T) = \frac{4}{3}\left(\frac{T}{\bar{T}}\right)^{1/3},
%\label{eq:f}
\end{align*}
such that 
$\theta(\mathbf{x},t) = F(T) = \left(T(\mathbf{x},t)^4/\bar{T}\right)^{1/3}$ 
and 
$T(\mathbf{x},t) = F^{-1}(\theta)= \left(\bar{T}\theta(\mathbf{x},t)^3\right)^{1/4}$.

%%%%%%%%%%%%%%%%%%%%%%%%%%%%%%%%%%%%%%%%%%%%%%%%%%%%%%%%%%%%%%%%%%%%%%%%%%%%
\begin{figure}%[!h]
\centering%\fbox{%
\includegraphics[scale=.31,clip=true,trim=1.5in 0in 0in 0in]{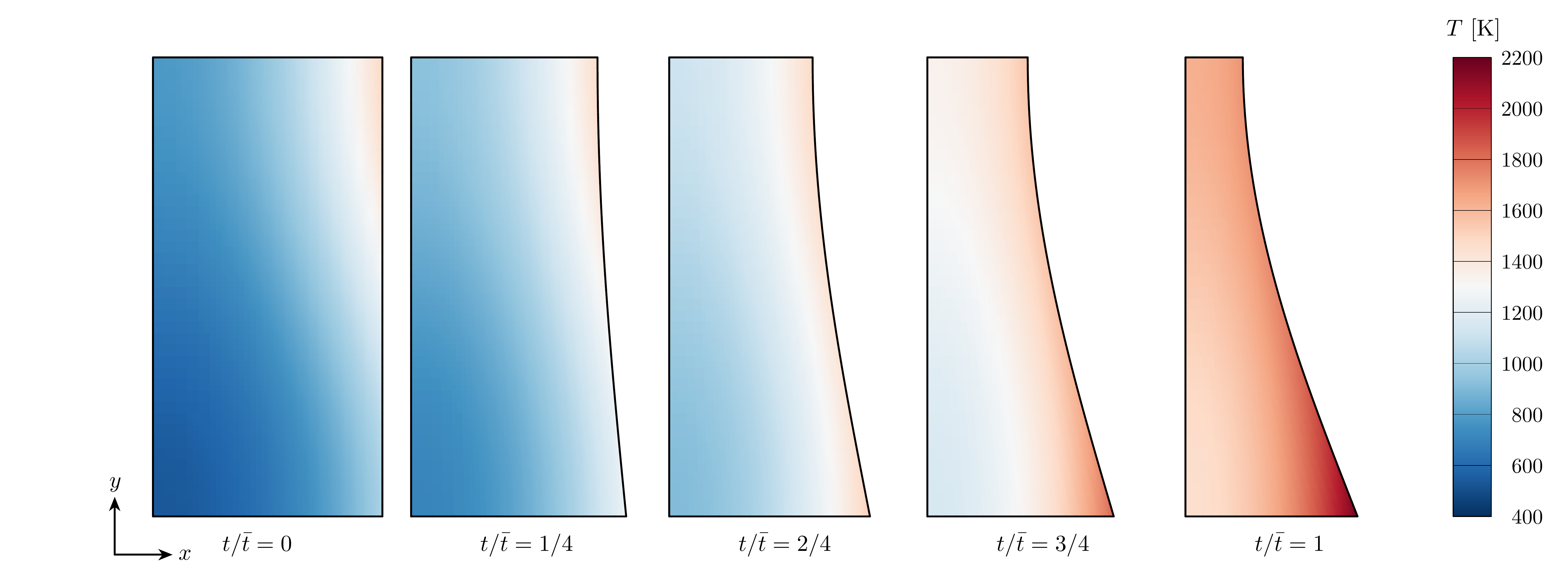}%}
\caption{Cartesian coordinates: Temperature at multiple instances in time for $\bar{\alpha}=10^{-5}$~m$^2$/s.}
\label{fig:cartesian_temperature}
\end{figure}

\begin{table}%[!h]
\centering
\begin{adjustbox}{max width=\textwidth}
\begin{tabular}{c c c c c c c}
\toprule
                         & \multicolumn{2}{c}{$T_s$ [$\times 10^3$ K]} & \multicolumn{2}{c}{$c_p(T_s)$ [J/kg/K]}       & \multicolumn{2}{c}{$k(T_s)$ [$\times 10^{-1}$ W/m/K]} \\
                           \cmidrule(r){2-3}                              \cmidrule(lr){4-5}                                \cmidrule(l){6-7}                                        
$\bar{\alpha}$ [m$^2$/s] & Min.             & Max.                     & Min.                  & Max.                  & Min.     & Max.                                       \\ \midrule
$10^{-8}$                & $0.7255$         & $1.4864$                 & $5.8150\times 10^{4}$ & $7.3855\times 10^{4}$ & $5.8150$ & $7.3855$                                   \\
$10^{-7}$                & $0.7322$         & $1.4864$                 & $5.8328\times 10^{3}$ & $7.3855\times 10^{3}$ & $5.8328$ & $7.3855$                                   \\
$10^{-6}$                & $0.8006$         & $1.4864$                 & $6.0089\times 10^{2}$ & $7.3855\times 10^{2}$ & $6.0089$ & $7.3855$                                   \\
$10^{-5}$                & $1.0133$         & $2.1674$                 & $6.5000\times 10^{1}$ & $8.3749\times 10^{1}$ & $6.5000$ & $8.3749$                                   \\
\bottomrule
\end{tabular}
\end{adjustbox}
\caption{Cartesian coordinates: Extrema of $T_s$, $c_p(T_s)$, and $k(T_s)$ for $t\in[0,\,\bar{t}]$.}
\label{tab:cartesian_extrema_1}
\end{table}

\begin{table}%[!h]
\centering
\begin{adjustbox}{max width=\textwidth}
\begin{tabular}{c c c c c c c c c}
\toprule
                         & \multicolumn{2}{c}{$h_s(T_s)$ [J/kg]}          & \multicolumn{2}{c}{$h_w(T_s,\reviewerTwo{p_e})$ [J/kg]} & \multicolumn{2}{c}{$B'(T_s,\reviewerTwo{p_e})$} & \multicolumn{2}{c}{$\reviewerTwo{C_e}(\xi,t)$ [kg/m$^2$/s]} \\
                           \cmidrule(l){2-3}                                \cmidrule(lr){4-5}                                        \cmidrule(lr){6-7}                                \cmidrule(l){8-9}
$\bar{\alpha}$ [m$^2$/s] & Min.                  & Max.                   & Min.                  & Max.                            & Min.     & Max.                                 & Min.     & Max.                                             \\ \midrule
$10^{-8}$                & $2.3040\times 10^{7}$ & $7.3732\times 10^{7}$  & $6.6657\times 10^{6}$ & $3.4600\times 10^{7}$           & $0.4071$ & $0.8842$                             & $0.9076$ & $3.3377$                                         \\
$10^{-7}$                & $2.3429\times 10^{6}$ & $7.3732\times 10^{6}$  & $6.8106\times 10^{5}$ & $3.4600\times 10^{6}$           & $0.4098$ & $0.8842$                             & $0.9076$ & $3.3213$                                         \\
$10^{-6}$                & $2.7476\times 10^{5}$ & $7.3732\times 10^{5}$  & $8.3796\times 10^{4}$ & $3.4600\times 10^{5}$           & $0.4388$ & $0.8842$                             & $0.9076$ & $3.1548$                                         \\
$10^{-5}$                & $4.0799\times 10^{4}$ & $1.2754\times 10^{5}$  & $1.4492\times 10^{4}$ & $8.0673\times 10^{4}$           & $0.5509$ & $1.7214$                             & $0.2704$ & $1.7339$                                         \\
\bottomrule
\end{tabular}
\end{adjustbox}
\caption{Cartesian coordinates: Extrema of $h_s(T_s)$, $h_w(T_s,\reviewerTwo{p_e})$, $B'(T_s,\reviewerTwo{p_e})$, and $\reviewerTwo{C_e}(\xi,t)$ for $t\in[0,\,\bar{t}]$.}
\label{tab:cartesian_extrema_2}
\end{table}

\begin{table}%[!h]
\centering
\begin{adjustbox}{max width=\textwidth}
\begin{tabular}{c c c c c}
\toprule
                         & \multicolumn{2}{c}{$h_r(\xi,t)$ [J/kg], \reviewerThree{$\epsilon\ne 0$}} & \multicolumn{2}{c}{$h_r(\xi,t)$ [J/kg], $\epsilon=0$} \\
                           \cmidrule(r){2-3}                                          \cmidrule(l){4-5}                           
$\bar{\alpha}$ [m$^2$/s] & Min.                  & Max.                             & Min.                  & Max.                          \\ \midrule
$10^{-8}$                & $1.2985\times 10^{4}$ & $2.1250\times 10^{5}$            & $6.0431\times 10^{3}$ & $7.3901\times 10^{4}$         \\
$10^{-7}$                & $1.3267\times 10^{4}$ & $2.1250\times 10^{5}$            & $6.1212\times 10^{3}$ & $7.3901\times 10^{4}$         \\
$10^{-6}$                & $1.6589\times 10^{4}$ & $2.1250\times 10^{5}$            & $6.9825\times 10^{3}$ & $7.3901\times 10^{4}$         \\
$10^{-5}$                & $9.6647\times 10^{4}$ & $4.7331\times 10^{6}$            & $3.8625\times 10^{4}$ & $5.6898\times 10^{5}$         \\
\bottomrule
\end{tabular}
\end{adjustbox}
\caption{Cartesian coordinates: Extrema of $h_r(\xi,t)$ for $t\in[0,\,\bar{t}]$.}
\label{tab:cartesian_extrema_3}
\end{table}
%%%%%%%%%%%%%%%%%%%%%%%%%%%%%%%%%%%%%%%%%%%%%%%%%%%%%%%%%%%%%%%%%%%%%%%%%%%%

%%%%%%%%%%%%%%%%%%%%%%%%%%%%%%%%%%%%%%%%%%%%%%%%%%%%%%%%%%%%%%%%%%%%%%%%%%%%
\begin{figure}%[th]
\centering
\begin{subfigure}[t]{.49\textwidth}
\includegraphics[scale=.64,clip=true,trim=2.25in 0in 2.8in 0in]{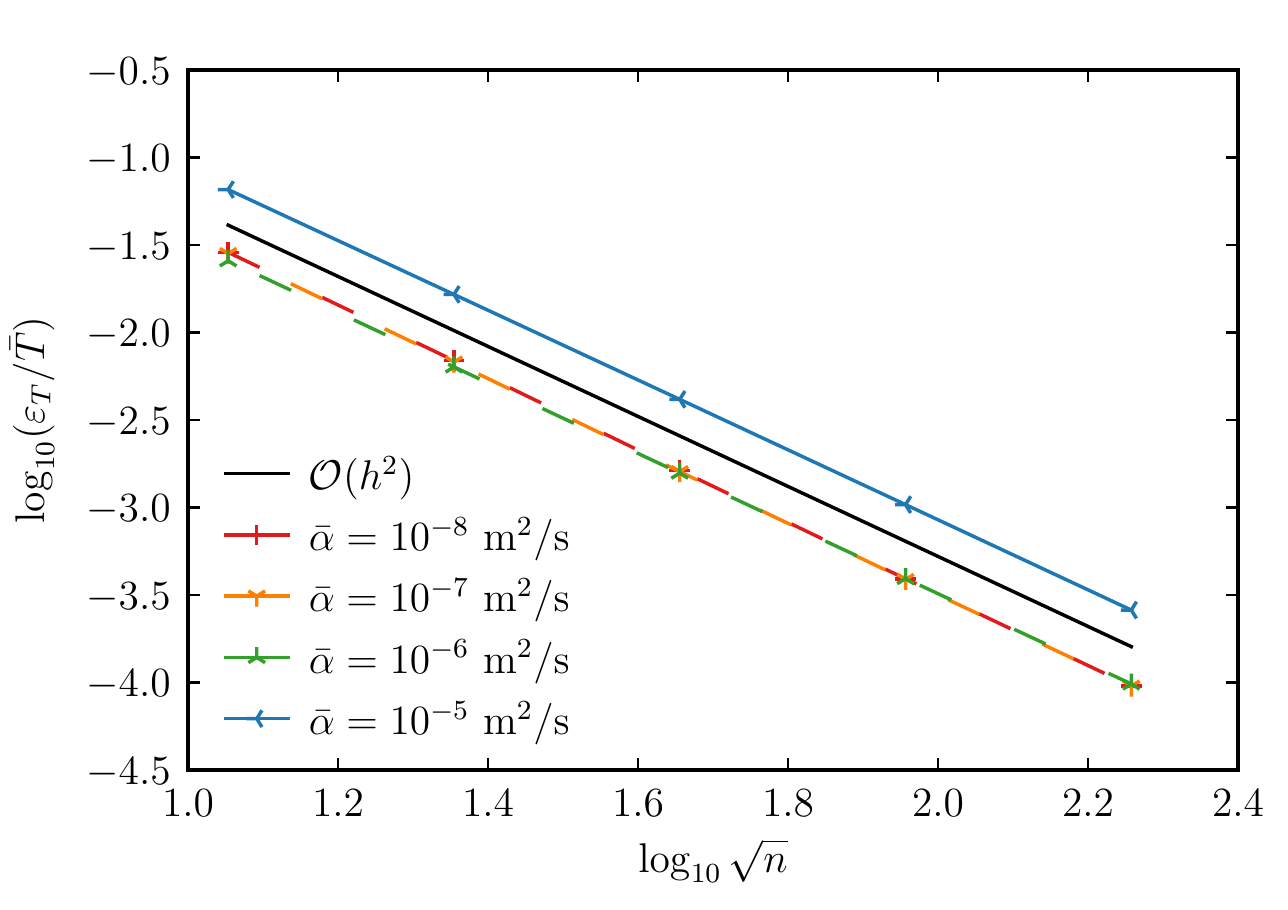}
\caption{\reviewerThree{$\epsilon\ne 0$}, $\Delta t/4$}
\label{fig:cartesian_T_so_rad}
\end{subfigure}
\hspace{0.25em}
\begin{subfigure}[t]{.49\textwidth}
\includegraphics[scale=.64,clip=true,trim=2.25in 0in 2.8in 0in]{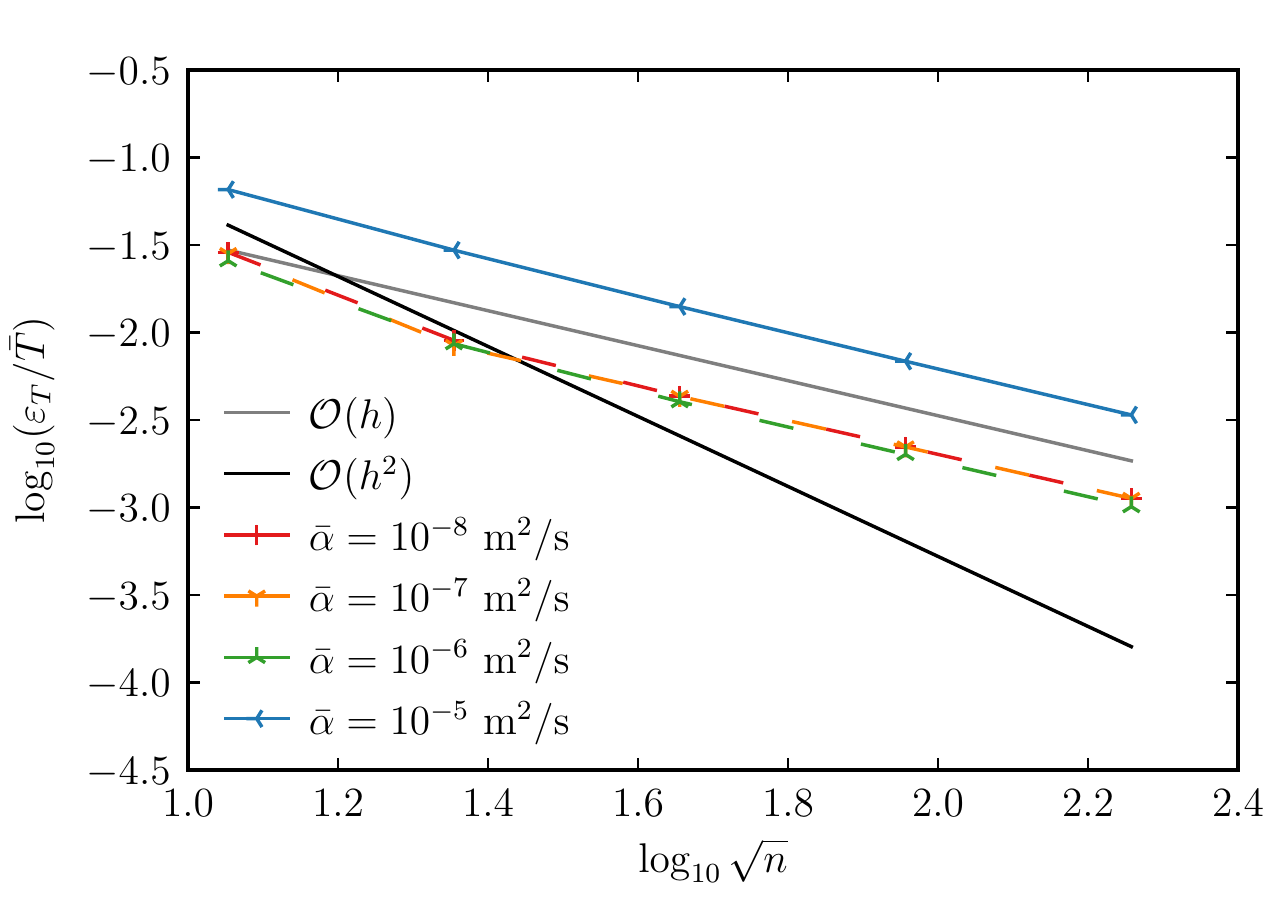}
\caption{\reviewerThree{$\epsilon\ne 0$}, \reviewerOneThree{$\Delta t/2$}}
\label{fig:cartesian_T_fo_rad}
\end{subfigure}
\\
\begin{subfigure}[t]{.49\textwidth}
\includegraphics[scale=.64,clip=true,trim=2.25in 0in 2.8in 0in]{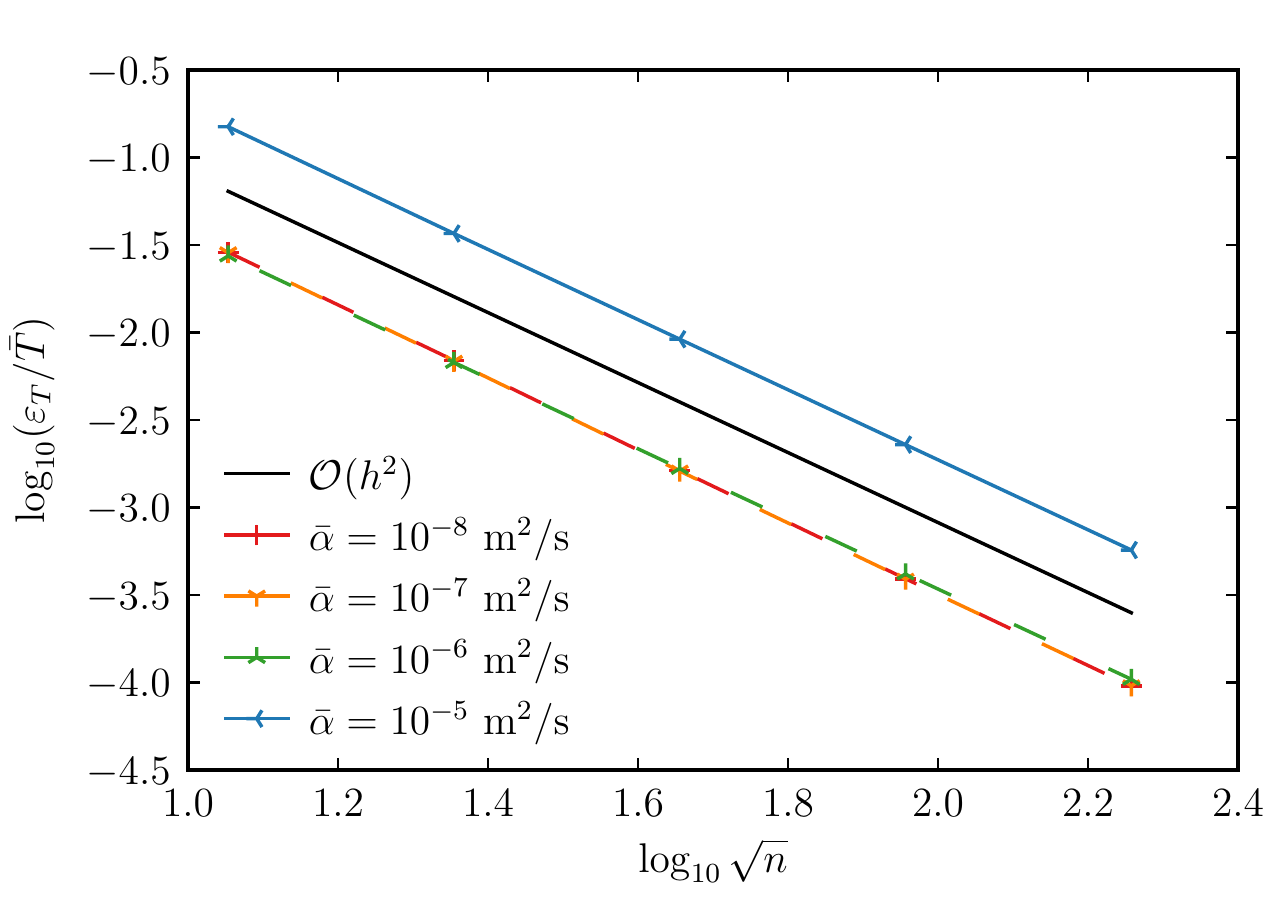}
\caption{$\epsilon=0$, $\Delta t/4$}
\label{fig:cartesian_T_so}
\end{subfigure}
\hspace{0.25em}
\begin{subfigure}[t]{.49\textwidth}
\includegraphics[scale=.64,clip=true,trim=2.25in 0in 2.8in 0in]{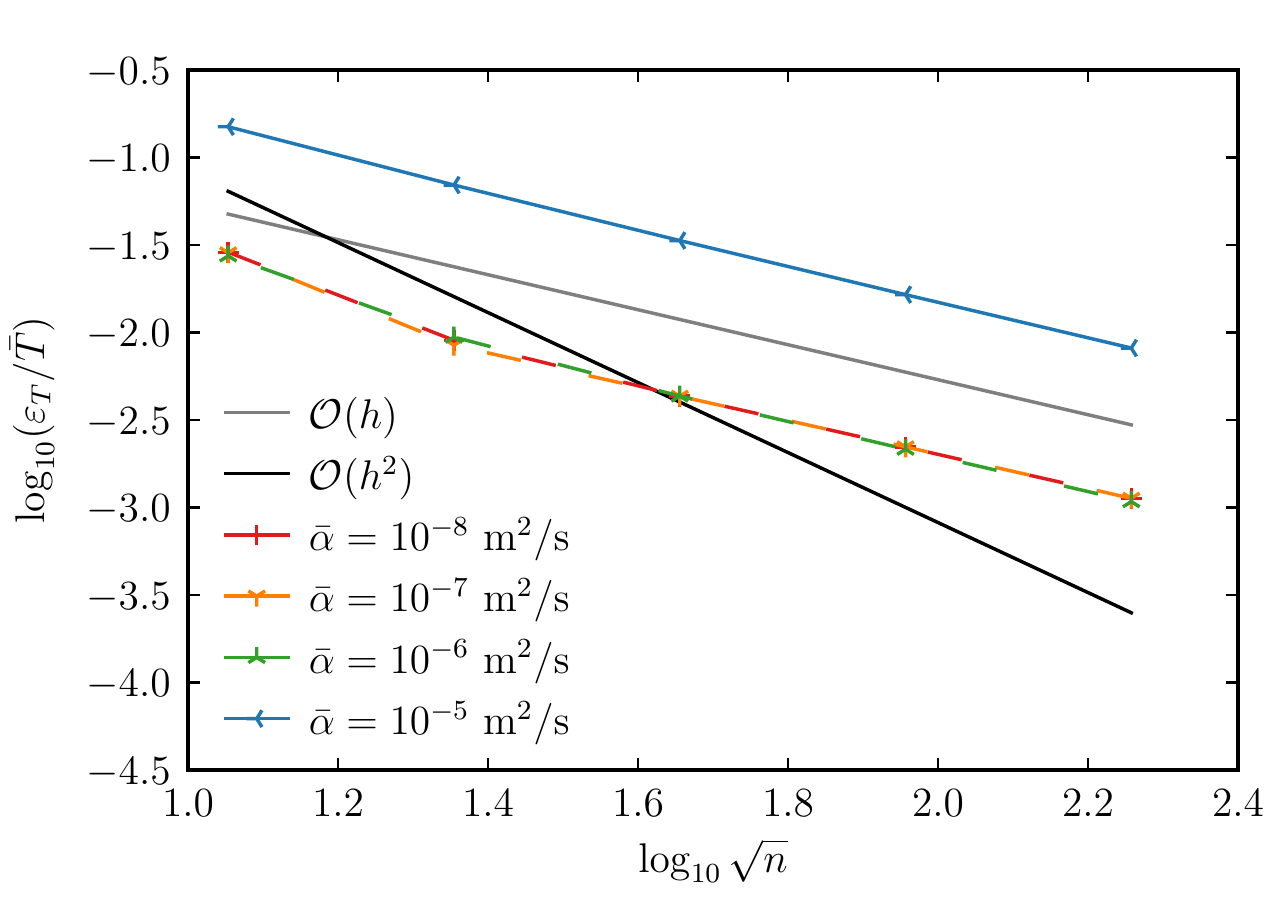}
\caption{$\epsilon=0$, \reviewerOneThree{$\Delta t/2$}}
\label{fig:cartesian_T_fo}
\end{subfigure}
\caption{Cartesian coordinates: Norm of the error for $T$.}
\vskip-\dp\strutbox
\label{fig:cartesian_T}
\end{figure}

\begin{table}%[bh]
\centering
\reviewerOneThree{%
\begin{tabular*}{.8\textwidth}{@{\extracolsep{\fill}} c c c c c c c c c c}
\toprule
& & \multicolumn{4}{c}{$\Delta t/4$, for $\bar{\alpha}$ [m$^2$/s]} & \multicolumn{4}{c}{$\Delta t/2$, for $\bar{\alpha}$ [m$^2$/s]} \\
\cmidrule(r){3-6} \cmidrule(l){7-10}
&
Mesh & 
$10^{-8}$ &
$10^{-7}$ &
$10^{-6}$ &
$10^{-5}$ &
$10^{-8}$ &
$10^{-7}$ &
$10^{-6}$ &
$10^{-5}$ \\ 
\midrule
\multirow{4}{*}{\rotatebox[origin=c]{90}{$\epsilon\ne 0$}}
& 1--2 & 2.0545 & 2.0588 & 2.0146 & 1.9901 & 1.6710 & 1.7444 & 1.5752 & 1.1496 \\
& 2--3 & 2.0832 & 2.0790 & 2.0152 & 1.9919 & 1.0589 & 0.9648 & 1.1002 & 1.0714 \\
& 3--4 & 2.0595 & 2.0492 & 2.0016 & 1.9979 & 0.9646 & 0.9602 & 1.0008 & 1.0372 \\
& 4--5 & 2.0361 & 2.0219 & 2.0015 & 2.0013 & 0.9755 & 0.9771 & 0.9919 & 1.0190 \\
\midrule
\multirow{4}{*}{\rotatebox[origin=c]{90}{$\epsilon=0$}}
& 1--2 & 2.0543 & 2.0600 & 2.0218 & 2.0265 & 1.6713 & 1.7545 & 1.5411 & 1.1061 \\
& 2--3 & 2.0838 & 2.0841 & 2.0122 & 2.0063 & 1.0601 & 0.9684 & 1.1190 & 1.0538 \\
& 3--4 & 2.0600 & 2.0531 & 2.0022 & 2.0017 & 0.9648 & 0.9604 & 1.0092 & 1.0272 \\
& 4--5 & 2.0366 & 2.0234 & 2.0001 & 2.0004 & 0.9755 & 0.9772 & 0.9951 & 1.0137 \\
\bottomrule
\end{tabular*}
}
\caption{Cartesian coordinates: Observed order of accuracy $p$ for $T$.}
\label{tab:cartesian_rates_T}
\end{table}
%%%%%%%%%%%%%%%%%%%%%%%%%%%%%%%%%%%%%%%%%%%%%%%%%%%%%%%%%%%%%%%%%%%%%%%%%%%%
\begin{figure}%[!h]
\centering
\begin{subfigure}[t]{.49\textwidth}
\includegraphics[scale=.64,clip=true,trim=2.25in 0in 2.8in 0in]{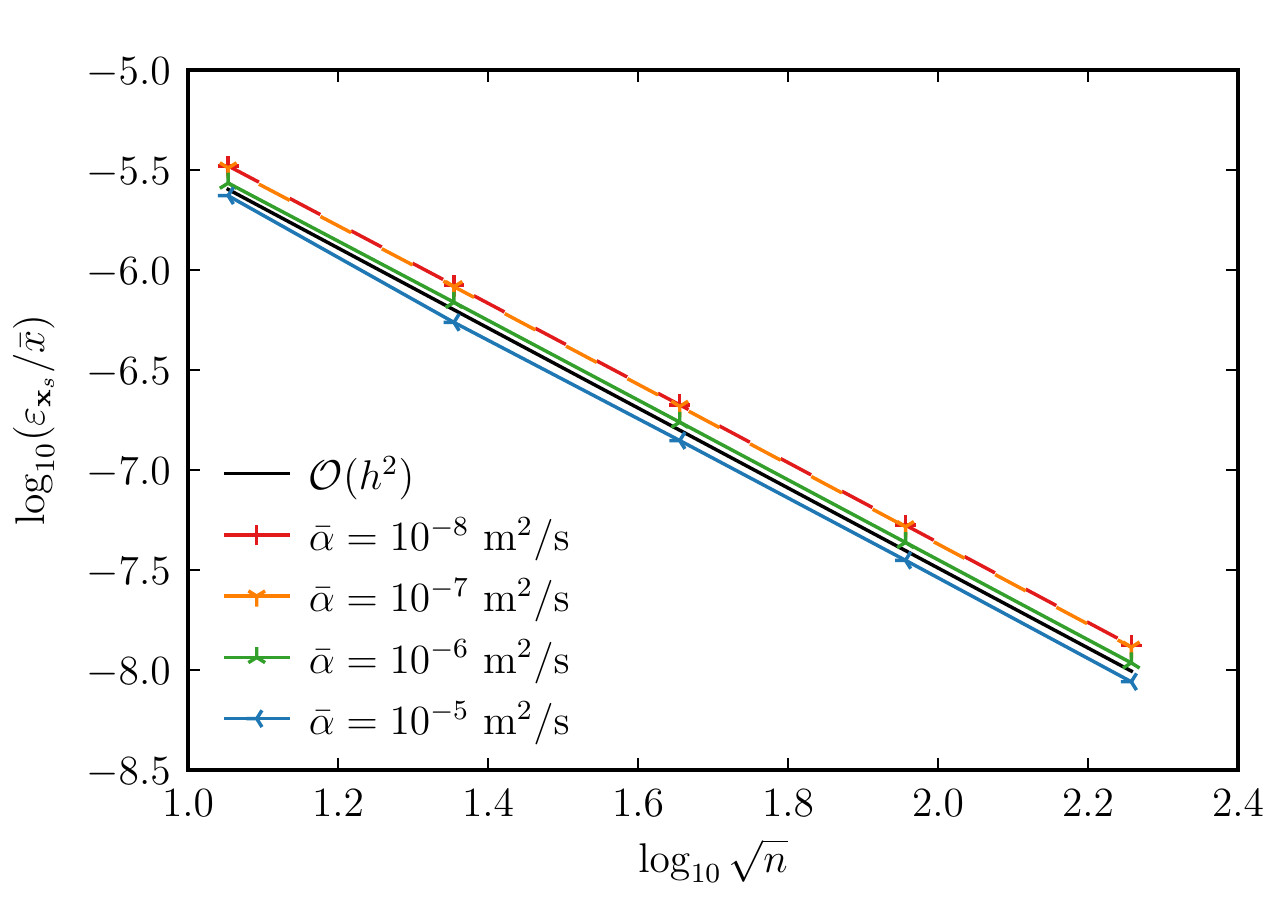}
\caption{\reviewerThree{$\epsilon\ne 0$}, $\Delta t/4$}
\label{fig:cartesian_xs_so_rad}
\end{subfigure}
\hspace{0.25em}
\begin{subfigure}[t]{.49\textwidth}
\includegraphics[scale=.64,clip=true,trim=2.25in 0in 2.8in 0in]{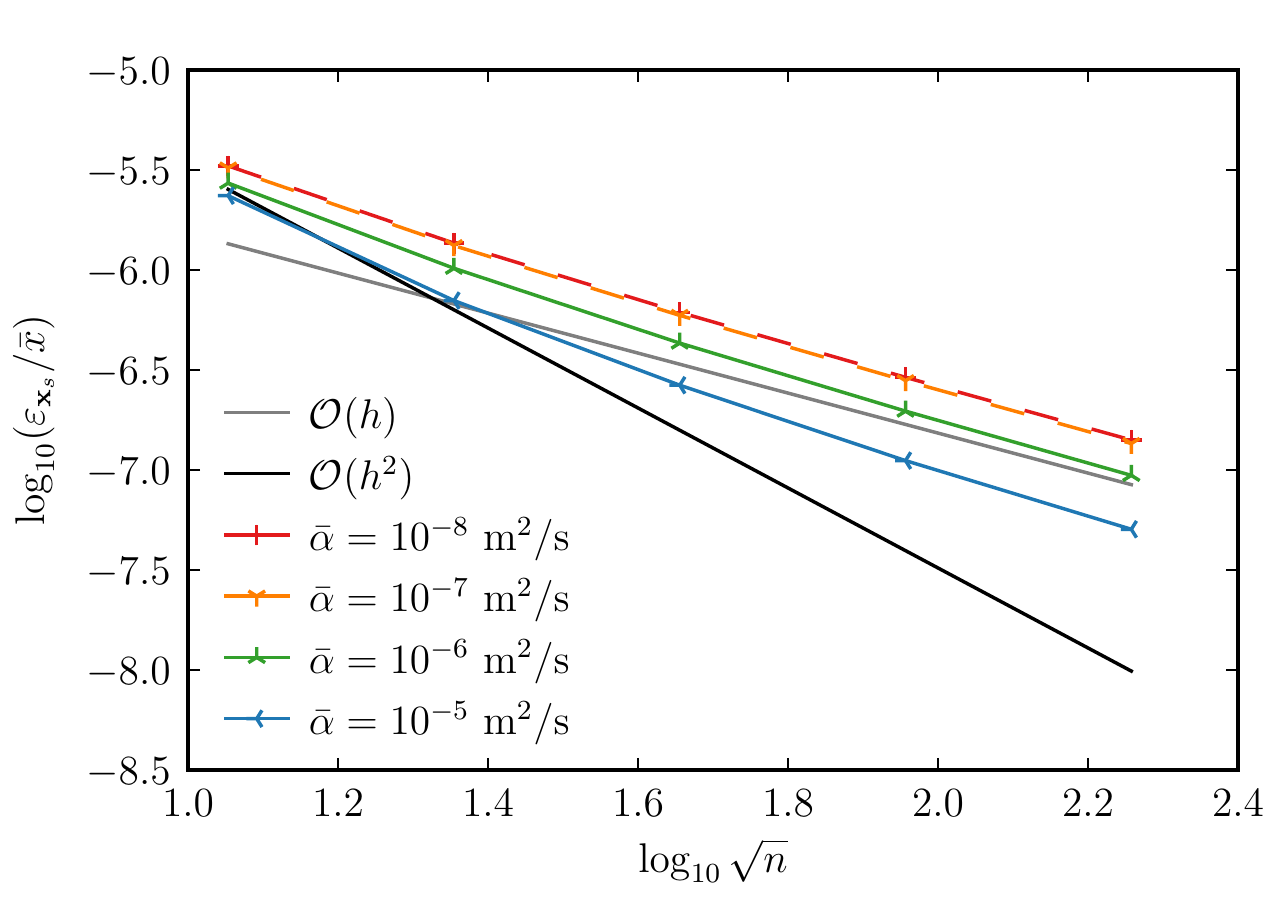}
\caption{\reviewerThree{$\epsilon\ne 0$}, \reviewerOneThree{$\Delta t/2$}}
\label{fig:cartesian_xs_fo_rad}
\end{subfigure}
\\
\begin{subfigure}[t]{.49\textwidth}
\includegraphics[scale=.64,clip=true,trim=2.25in 0in 2.8in 0in]{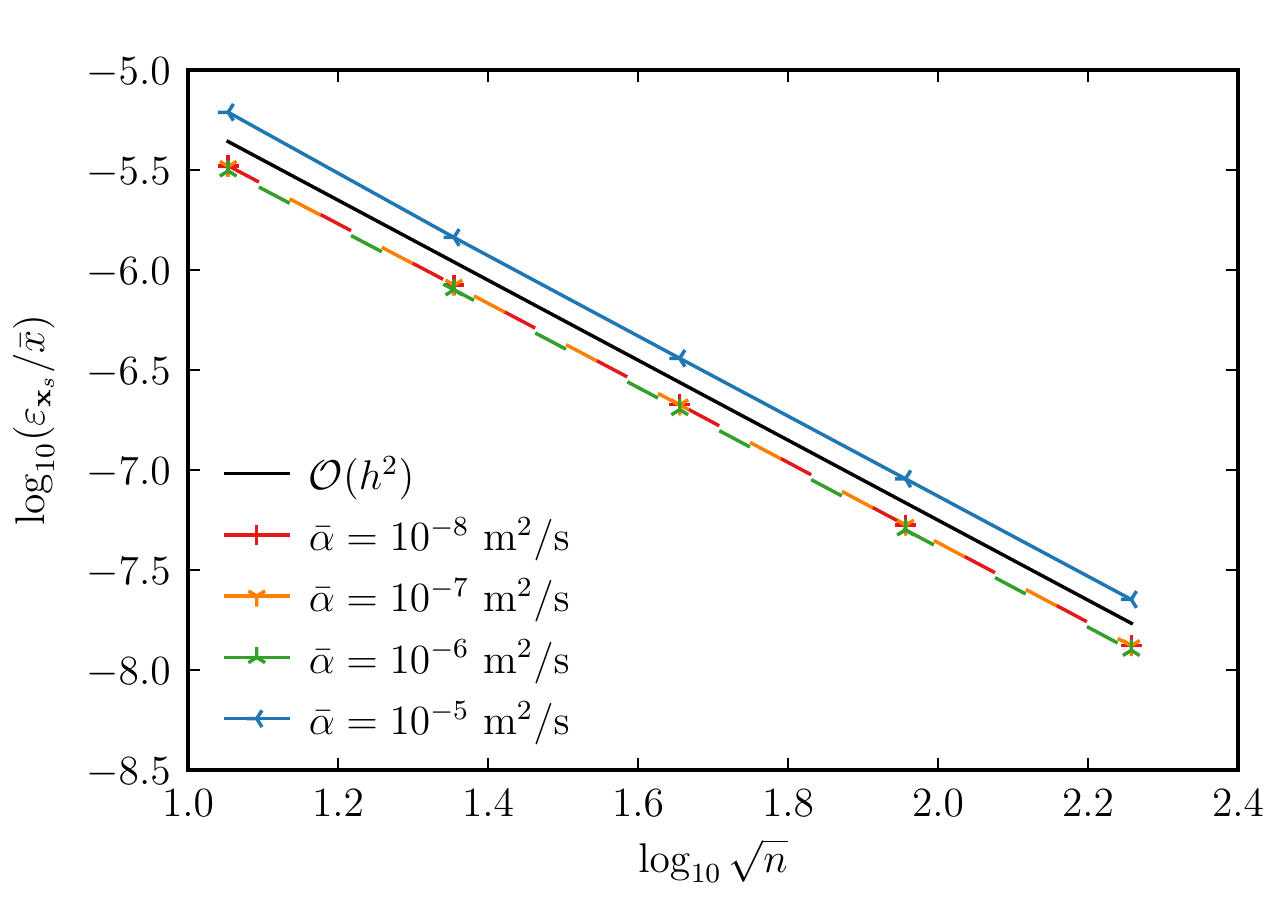}
\caption{$\epsilon=0$, $\Delta t/4$}
\label{fig:cartesian_xs_so}
\end{subfigure}
\hspace{0.25em}
\begin{subfigure}[t]{.49\textwidth}
\includegraphics[scale=.64,clip=true,trim=2.25in 0in 2.8in 0in]{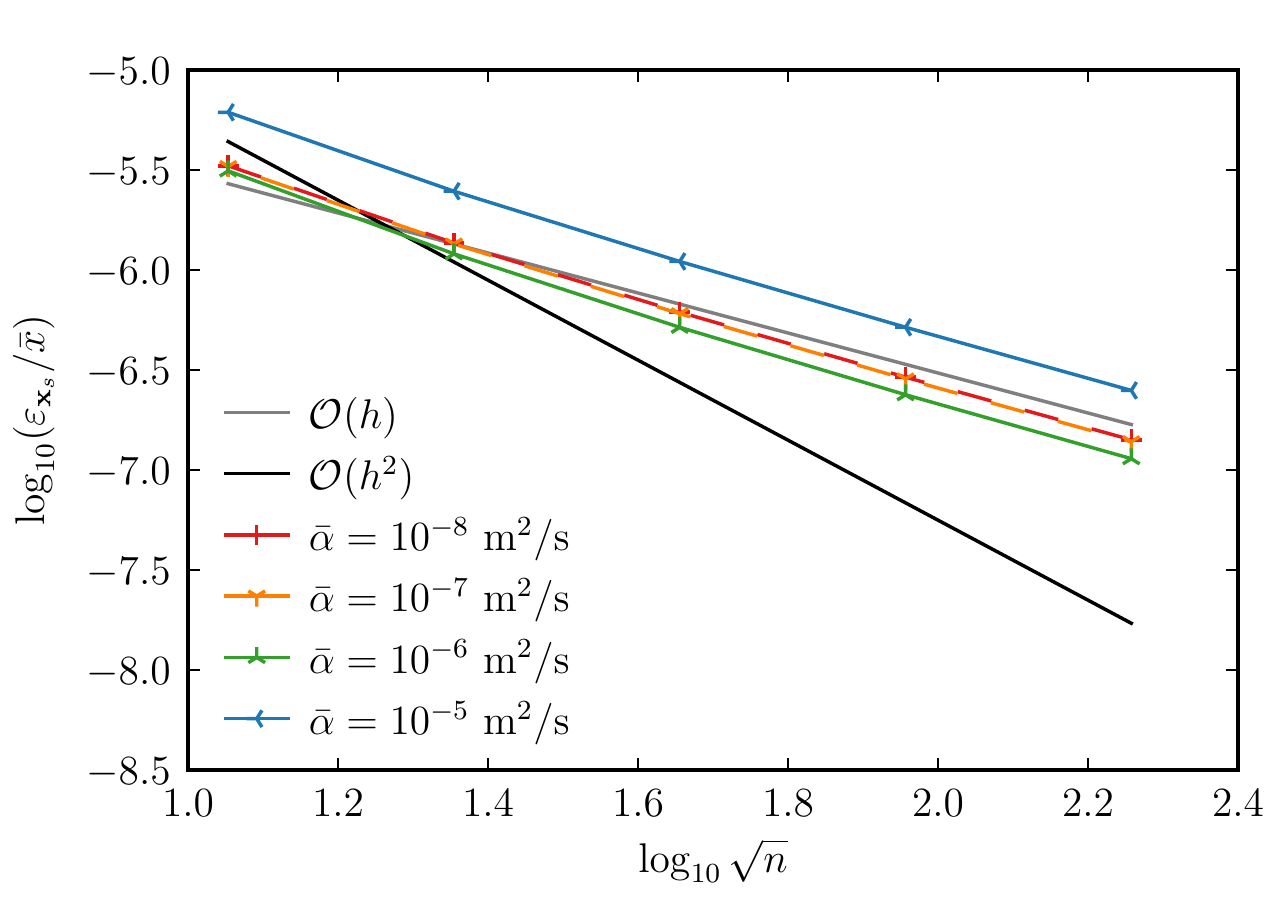}
\caption{$\epsilon=0$, \reviewerOneThree{$\Delta t/2$}}
\label{fig:cartesian_xs_fo}
\end{subfigure}
\caption{Cartesian coordinates: Norm of the error for $\mathbf{x}_s$.}
\vskip-\dp\strutbox
\label{fig:cartesian_xs}
\end{figure}

%%%%%%%%%%%%%%%%%
\begin{table}
\centering
\reviewerOneThree{%
\begin{tabular*}{.8\textwidth}{@{\extracolsep{\fill}} c c c c c c c c c c}
\toprule
& & \multicolumn{4}{c}{$\Delta t/4$, for $\bar{\alpha}$ [m$^2$/s]} & \multicolumn{4}{c}{$\Delta t/2$, for $\bar{\alpha}$ [m$^2$/s]} \\
\cmidrule(r){3-6} \cmidrule(l){7-10}
&
Mesh & 
$10^{-8}$ &
$10^{-7}$ &
$10^{-6}$ &
$10^{-5}$ &
$10^{-8}$ &
$10^{-7}$ &
$10^{-6}$ &
$10^{-5}$ \\ 
\midrule
\multirow{4}{*}{\rotatebox[origin=c]{90}{$\epsilon\ne 0$}}
& 1--2 & 1.9809 & 1.9715 & 1.9796 & 2.1040 & 1.2825 & 1.2891 & 1.4152 & 1.7413 \\
& 2--3 & 1.9876 & 1.9883 & 1.9914 & 1.9620 & 1.1490 & 1.1591 & 1.2422 & 1.4057 \\
& 3--4 & 1.9983 & 1.9992 & 1.9975 & 1.9898 & 1.0799 & 1.0854 & 1.1302 & 1.2524 \\
& 4--5 & 1.9991 & 2.0002 & 2.0001 & 2.0169 & 1.0407 & 1.0435 & 1.0667 & 1.1405 \\
\midrule
\multirow{4}{*}{\rotatebox[origin=c]{90}{$\epsilon=0$}}
& 1--2 & 1.9809 & 1.9712 & 1.9744 & 2.0785 & 1.2822 & 1.2862 & 1.3775 & 1.3111 \\
& 2--3 & 1.9876 & 1.9881 & 1.9906 & 2.0079 & 1.1489 & 1.1572 & 1.2206 & 1.1667 \\
& 3--4 & 1.9983 & 1.9992 & 1.9974 & 2.0016 & 1.0798 & 1.0845 & 1.1198 & 1.0927 \\
& 4--5 & 1.9991 & 2.0001 & 2.0000 & 2.0037 & 1.0407 & 1.0431 & 1.0620 & 1.0491 \\
\bottomrule
\end{tabular*}
}
\caption{Cartesian coordinates: Observed order of accuracy $p$ for $\mathbf{x}_s$.}
\label{tab:cartesian_rates_xs}
\end{table}

%%%%%%%%%%%%%%%%%%%%%%%%%%%%%%%%%%%%%%%%%%%%%%%%%%%%%%%%%%%%%%%%%%%%%%%%%%%%

For $\theta(\mathbf{x},t)$, we truncate~\eqref{eq:full_theta_form} to $\max{i}=0$ and $\max{j}=1$.  
We choose this truncation because, \reviewerThree{from~\eqref{eq:v_j}}, $v_0(y)=1$ and $v_1(y)=\cos(\pi y/H)$  enable us to obtain variation with respect to $y$ without $\theta(\mathbf{x},t)$ becoming negative.  For $u_0(x)$, we choose an imaginary valued $\mu_0 = 3\sqrt{-1}/(2 W)$, such that $\lambda_{i,j}$ \reviewerThree{from}~\eqref{eq:cartesian_lambda} is negative and the temperature increases with time.  $u_0(x)=\cosh(3x/(2W))$ \reviewerThree{from}~\eqref{eq:u_i} then provides sufficient variation with respect to $x$.
We set $\hat{\theta}_{{0,0}_0} = 400$~K, and $\hat{\theta}_{{0,1}_0} = -100$~K.  With these choices,~\eqref{eq:full_theta_form} becomes
\begin{align*}
\theta(\mathbf{x},t) =  100 e^{22,500\bar{\alpha} t} \big(4 - e^{-2500\pi^2\bar{\alpha} t}\cos(\pi y/H)\big)\cosh(3 x/(2W))\text{ K}.%\sum_{i,j=0}^\infty\hat{\theta}_{i,j}(t) \varphi_{i,j}(\mathbf{x}),
\end{align*}

To manufacture the recession, we manufacture
\begin{align*}
\mathbf{x}_s(\xi,t) = \left\{
%\begin{matrix}\displaystyle 
W\left(1- \frac{t}{\bar{t}}\frac{1+2\sin\left(\pi\xi/2\right)}{4}\right),\,  H\xi
%\end{matrix}
\right\},
%\label{eq:x_s}
\end{align*}
which has the initial condition $\mathbf{x}_s(\xi,0)=\{W,\xi H\}$, such that the initial domain is a rectangle.  $\xi$ is related to $\mathbf{x}_s$ by $\xi=y_s/H$.
We set $W = 1$~cm and $H = 2$~cm.
Figure~\ref{fig:cartesian_domain} shows the evolution of the domain, and Figure~\ref{fig:cartesian_temperature} shows the evolution of $T$ for $\bar{\alpha}=10^{-5}$~m$^2$/s.  \reviewerThree{Tables~\ref{tab:cartesian_extrema_1} and~\ref{tab:cartesian_extrema_2} list the extrema of $T_s$, $c_p(T_s)$, $k(T_s)$, $h_s(T_s)$, $h_w(T_s,\reviewerTwo{p_e})$, $B'(T_s,\reviewerTwo{p_e})$, and $\reviewerTwo{C_e}(\xi,t)$ for $t\in[0,\,\bar{t}]$.  Table~\ref{tab:cartesian_extrema_3} lists the extrema of $h_r(\xi,t)$ with ($\epsilon\ne 0$) and without ($\epsilon=0$) the radiative flux}.  These tables provide a brief summary of the magnitudes of these quantities.  
\reviewerThree{%
Even though $T(\mathbf{x},t)$ increases with time for a given $\mathbf{x}$, $T_s$ may not necessarily do so, due to the motion of $\mathbf{x}_s$.  For $\bar{\alpha}=10^{-5}$~m$^2$/s, the maximum value of $T_s$ occurs when $t=\bar{t}$; for the other values of $\bar{\alpha}$, the maximum occurs when $t=0$.  Therefore, for the variables in Tables~\ref{tab:cartesian_extrema_1}--\ref{tab:cartesian_extrema_3}, one of the extrema is either constant or inversely proportional to $\bar{\alpha}$ for $\bar{\alpha}=\{10^{-8},\allowbreak 10^{-7},\allowbreak 10^{-6}\}$~m$^2$/s.
}

\reviewerOneThree{%
We consider five discretizations, for which we double the number of elements in both spatial dimensions for each subsequent discretization.  The coarsest discretization contains $8\times 16$ elements, and the finest contains $128\times 256$ elements.  For the time discretization, we quarter the time step ($\Delta t/4$) and halve the time step ($\Delta t/2$) for each subsequent discretization.  The coarsest discretization has a time step of 0.2~s, and the finest has a time step of 0.78125~ms for $\Delta t/4$ and 12.5~ms for $\Delta t/2$.  Because the spatial discretization is second-order accurate and the time-integration scheme is first-order accurate, we expect to achieve $\mathcal{O}(h^2)$ for $\Delta t/4$ and $\mathcal{O}(h)$ for $\Delta t/2$.  We additionally consider cases with and without the radiative flux.
}

\reviewerOneThree{%
For each of the five values of $\bar{\alpha}$ with ($\epsilon\ne 0$) and without ($\epsilon=0$) the radiative flux, Figures~\ref{fig:cartesian_T} and~\ref{fig:cartesian_xs} show how the error norms $\varepsilon_T$~\eqref{eq:error_norm_T} and $\varepsilon_{\mathbf{x}_s}$~\eqref{eq:error_norm_xs}, which are nondimensionalized by $\bar{T} = 1$~K and $\bar{x}=1$~m, vary with respect to $n$, which is the number of elements.  Additionally, Tables~\ref{tab:cartesian_rates_T} and~\ref{tab:cartesian_rates_xs} provide the observed order of accuracy $p$ between discretization pairs.  The error norms in both plots and tables are $\mathcal{O}(h^2)$ for $\Delta t/4$ and $\mathcal{O}(h)$ for $\Delta t/2$, as expected.
}

%-------------------------------------------------------------------------------
\subsection{Polar Coordinates} %------------------------------------------------
%-------------------------------------------------------------------------------

The second problem set we consider uses polar coordinates.  For this problem set, we use constant coefficients, such that $f(T) = 1$, and $T(\mathbf{x},t) = \theta(\mathbf{x},t)$.

For $\theta(\mathbf{x},t)$, we truncate~\eqref{eq:full_theta_form} to $\max{i}=0$ and $\max{j}=1$.  
We choose this truncation because, \reviewerThree{from~\eqref{eq:v_j_polar},} $v_0(\phi)=1$ and $v_1(\phi)=\cos(\pi\phi/\bar{\phi})$ enable us to obtain variation with respect to $\phi$ without $\theta(\mathbf{x},t)$ becoming negative.  For $u_{0,0}(r)$ and $u_{0,1}(r)$ \reviewerThree{in}~\eqref{eq:u_in}, we choose $\lambda_{0,0}=\lambda_{0,1}=-$22,500~m$^{-2}$, so that the temperature increases with time.  $u_{0,0}(r)$ and $u_{0,1}(r)$ provide sufficient variation with respect to $r$.
We set $\hat{\theta}_{{0,0}_0} = 200$~K, and $\hat{\theta}_{{0,1}_0} = 30$~K. 

\begin{figure}
\centering%\fbox{%
\includegraphics[scale=.31,clip=true,trim=1.5in 0in 0in 0in]{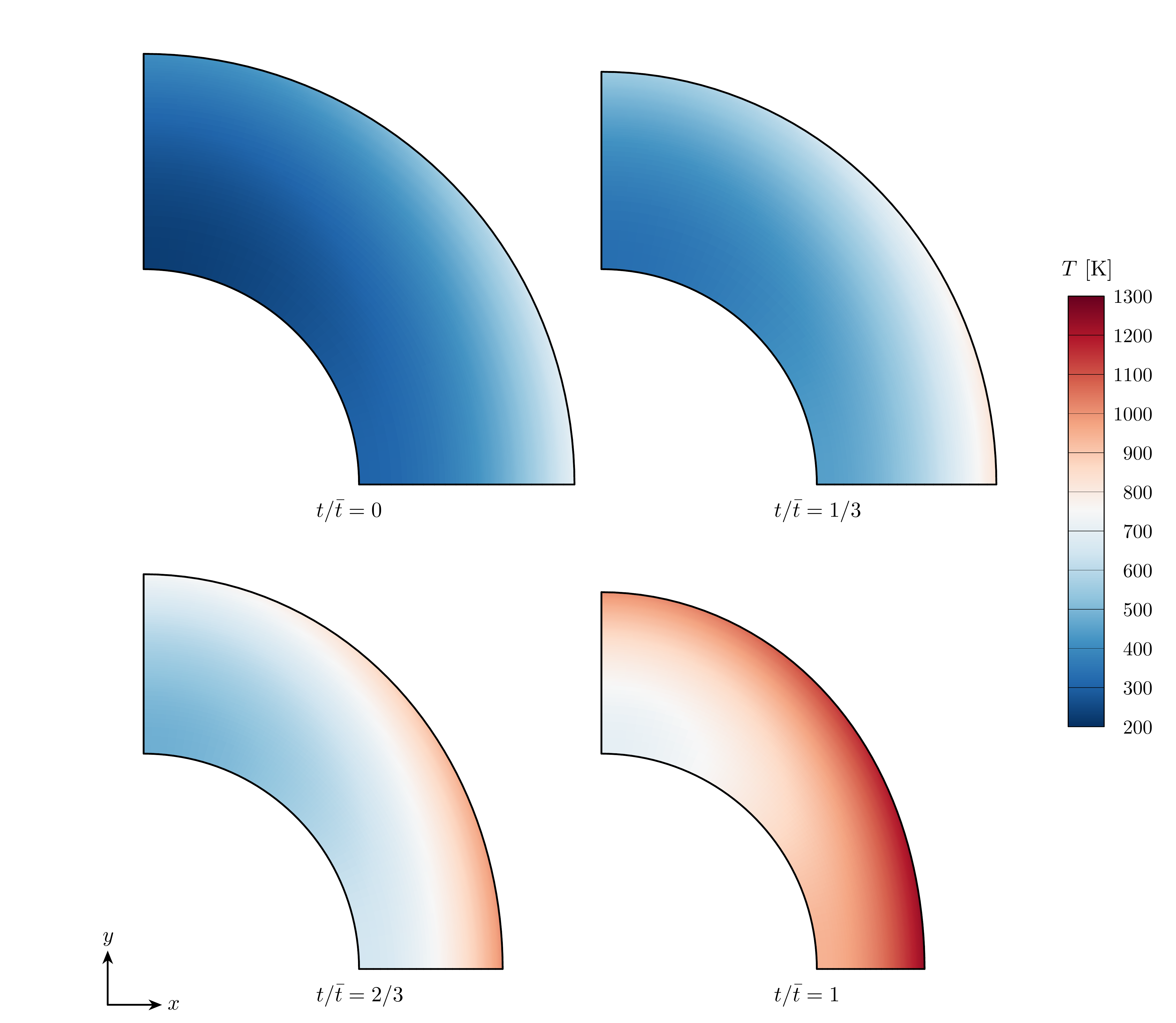}%} .31
\caption{Polar coordinates: Temperature at multiple instances in time for $\bar{\alpha}=10^{-5}$~m$^2$/s.}
\label{fig:polar_temperature}
\end{figure}

To manufacture the recession, we relate $\xi$ to $\mathbf{x}_s$ by $\xi=\phi_s/\bar{\phi}$, and we manufacture
\begin{align}
\mathbf{x}_s(\xi,t) = r_s(\xi,t)\left\{\cos\phi_s,\, \sin\phi_s\right\},
\label{eq:x_s_polar}
\end{align}
where
\begin{align*}
r_s(\xi,t) = r_1 - (r_1-r_0) \frac{t}{\bar{t}}\frac{3+\cos\left(\pi\xi\right)}{8}.
\end{align*}
Equation~\eqref{eq:x_s_polar} has the initial condition $\mathbf{x}_s(\xi,0)=r_1\{\cos\phi_s,\, \sin\phi_s\}$, such that the initial domain is a fractional annulus.  
%
%For the spatial domain, we set $r_0 = 1$~cm, $r_1 = 2$~cm, and $\bar{\phi}=\pi/2$, and, for the time domain, we set $\bar{t} = 5$~s.  
We set $r_0 = 1$~cm, $r_1 = 2$~cm, and $\bar{\phi}=\pi/2$.
Figure~\ref{fig:polar_domain} shows the evolution of the domain, and Figure~\ref{fig:polar_temperature} shows the evolution of $T$ for $\bar{\alpha}=10^{-5}$~m$^2$/s.
\reviewerThree{%
Tables~\ref{tab:polar_extrema_1} and~\ref{tab:polar_extrema_2} list the extrema of $T_s$, $c_p(T_s)$, $k(T_s)$, $h_s(T_s)$, $h_w(T_s,\reviewerTwo{p_e})$, $B'(T_s,\reviewerTwo{p_e})$, and $\reviewerTwo{C_e}(\xi,t)$ for $t\in[0,\,\bar{t}]$.  Table~\ref{tab:polar_extrema_3} lists the extrema of $h_r(\xi,t)$ with ($\epsilon\ne 0$) and without ($\epsilon=0$) the radiative flux}.  These tables provide a brief summary of the magnitudes of these quantities.  \reviewerThree{For $\bar{\alpha}=10^{-5}$~m$^2$/s, the maximum value of $T_s$ occurs when $t=\bar{t}$; for the other values of $\bar{\alpha}$, the maximum occurs when $t=0$.  Therefore, for the variables in Tables~\ref{tab:polar_extrema_1}--\ref{tab:polar_extrema_3}, one of the extrema is either constant or inversely proportional to $\bar{\alpha}$ for $\bar{\alpha}=\{10^{-8},\allowbreak 10^{-7},\allowbreak 10^{-6}\}$~m$^2$/s.}

%%%%%%%%%%%%%%%%%%%%%%%%%%%%%%%%%%%%%%%%%%%%%%%%%%%%%%%%%%%%%%%%%%%%%%%%%%%%
%%%%%%%%%%%%%%%%%%%%%%%%%%%%%%%%%%%%%%%%%%%%%%%%%%%%%%%%%%%%%%%%%%%%%%%%%%%%
%%%%%%%%%%%%%%%%%%%%%%%%%%%%%%%%%%%%%%%%%%%%%%%%%%%%%%%%%%%%%%%%%%%%%%%%%%%%
\begin{table}%[t]
\centering
\begin{adjustbox}{max width=\textwidth}
\begin{tabular}{c c c c c c c}
\toprule
                         & \multicolumn{2}{c}{$T_s$ [$\times 10^3$ K]} & \multicolumn{2}{c}{$c_p(T_s)$ [J/kg/K]}       & \multicolumn{2}{c}{$k(T_s)$ [$\times 10^{-1}$ W/m/K]} \\
                           \cmidrule(r){2-3}                              \cmidrule(lr){4-5}                                \cmidrule(l){6-7}                                       
$\bar{\alpha}$ [m$^2$/s] & Min.             & Max.                     & Min.                  & Max.                  & Min.     & Max.                                       \\ \midrule
$10^{-8}$                & $0.3268$         & $0.6994$                 & $7.0000\times 10^{4}$ & $7.0000\times 10^{4}$ & $7.0000$ & $7.0000$                                   \\
$10^{-7}$                & $0.3301$         & $0.6994$                 & $7.0000\times 10^{3}$ & $7.0000\times 10^{3}$ & $7.0000$ & $7.0000$                                   \\
$10^{-6}$                & $0.3653$         & $0.6994$                 & $7.0000\times 10^{2}$ & $7.0000\times 10^{2}$ & $7.0000$ & $7.0000$                                   \\
$10^{-5}$                & $0.4110$         & $1.2353$                 & $7.0000\times 10^{1}$ & $7.0000\times 10^{1}$ & $7.0000$ & $7.0000$                                   \\
\bottomrule
\end{tabular}
\end{adjustbox}
\caption{Polar coordinates: Extrema of $T_s$, $c_p(T_s)$, and $k(T_s)$ for $t\in[0,\,\bar{t}]$.}
\label{tab:polar_extrema_1}
\end{table}

\begin{table}%[t]
\centering
\begin{adjustbox}{max width=\textwidth}
\begin{tabular}{c c c c c c c c c}
\toprule
                         & \multicolumn{2}{c}{$h_s(T_s)$ [J/kg]}         & \multicolumn{2}{c}{$h_w(T_s,\reviewerTwo{p_e})$ [J/kg]} & \multicolumn{2}{c}{$B'(T_s,\reviewerTwo{p_e})$ [$\times 10^{-1}$]} & \multicolumn{2}{c}{$\reviewerTwo{C_e}(\xi,t)$ [kg/m$^2$/s]} \\
                           \cmidrule(r){2-3}                               \cmidrule(lr){4-5}                                        \cmidrule(lr){6-7}                                                   \cmidrule(l){8-9}
$\bar{\alpha}$ [m$^2$/s] & Min.                  & Max.                  & Min.                  & Max.                            & Min.     & Max.                                                    & Min.     & Max.                                             \\ \midrule
$10^{-8}$                & $3.7530\times 10^{6}$ & $2.9834\times 10^{7}$ & $8.0534\times 10^{5}$ & $8.5613\times 10^{6}$           & $2.7321$ & $4.0245$                                                & $1.6576$ & $3.3967$                                         \\
$10^{-7}$                & $3.9858\times 10^{5}$ & $2.9834\times 10^{6}$ & $8.5752\times 10^{4}$ & $8.5613\times 10^{5}$           & $2.7412$ & $4.0245$                                                & $1.6576$ & $3.3828$                                         \\
$10^{-6}$                & $6.4478\times 10^{4}$ & $2.9834\times 10^{5}$ & $1.4259\times 10^{4}$ & $8.5613\times 10^{4}$           & $2.8393$ & $4.0245$                                                & $1.6576$ & $3.2397$                                         \\
$10^{-5}$                & $9.6514\times 10^{3}$ & $6.7348\times 10^{4}$ & $2.2366\times 10^{3}$ & $2.7206\times 10^{4}$           & $3.0164$ & $6.7772$                                                & $0.9285$ & $2.4848$                                         \\
\bottomrule
\end{tabular}
\end{adjustbox}
\caption{Polar coordinates: Extrema of $h_s(T_s)$, $h_w(T_s,\reviewerTwo{p_e})$, $B'(T_s,\reviewerTwo{p_e})$, and $\reviewerTwo{C_e}(\xi,t)$ for $t\in[0,\,\bar{t}]$.}
\label{tab:polar_extrema_2}
\end{table}

\begin{table}%[t]
\centering
\begin{adjustbox}{max width=\textwidth}
\begin{tabular}{c c c c c}
\toprule
                         & \multicolumn{2}{c}{$h_r(\xi,t)$ [J/kg], \reviewerThree{$\epsilon\ne 0$}} & \multicolumn{2}{c}{$h_r(\xi,t)$ [J/kg], $\epsilon=0$} \\
                           \cmidrule(r){2-3}                                          \cmidrule(l){4-5}                           
$\bar{\alpha}$ [m$^2$/s] & Min.                  & Max.                             & Min.                  & Max.                          \\ \midrule
$10^{-8}$                & $7.9476\times 10^{3}$ & $2.9082\times 10^{4}$            & $7.6789\times 10^{3}$ & $2.4335\times 10^{4}$         \\
$10^{-7}$                & $8.0748\times 10^{3}$ & $2.9082\times 10^{4}$            & $7.7888\times 10^{3}$ & $2.4335\times 10^{4}$         \\
$10^{-6}$                & $9.5107\times 10^{3}$ & $2.9082\times 10^{4}$            & $8.9993\times 10^{3}$ & $2.4335\times 10^{4}$         \\
$10^{-5}$                & $1.7725\times 10^{4}$ & $1.4163\times 10^{5}$            & $1.7095\times 10^{4}$ & $6.6926\times 10^{4}$         \\
\bottomrule
\end{tabular}
\end{adjustbox}
\caption{Polar coordinates: Extrema of $h_r(\xi,t)$ for $t\in[0,\,\bar{t}]$.}
\label{tab:polar_extrema_3}
\end{table}
%%%%%%%%%%%%%%%%%%%%%%%%%%%%%%%%%%%%%%%%%%%%%%%%%%%%%%%%%%%%%%%%%%%%%%%%%%%%

\reviewerOneThree{%
We consider five discretizations, for which we double the number of elements in both spatial dimensions for each subsequent discretization.  The coarsest discretization contains $8\times 18$ elements, and the finest contains $128\times 288$ elements.  For the time discretization, we quarter the time step ($\Delta t/4$) and halve the time step ($\Delta t/2$) for each subsequent discretization.  The coarsest discretization has a time step of 0.2~s, and the finest has a time step of 0.78125~ms for $\Delta t/4$ and 12.5~ms for $\Delta t/2$.  Because the spatial discretization is second-order accurate and the time-integration scheme is first-order accurate, we expect to achieve $\mathcal{O}(h^2)$ for $\Delta t/4$ and $\mathcal{O}(h)$ for $\Delta t/2$.  We additionally consider cases with and without the radiative flux.
}

\begin{figure}%[!h]
\centering
\begin{subfigure}[t]{.49\textwidth}
\includegraphics[scale=.64,clip=true,trim=2.25in 0in 2.8in 0in]{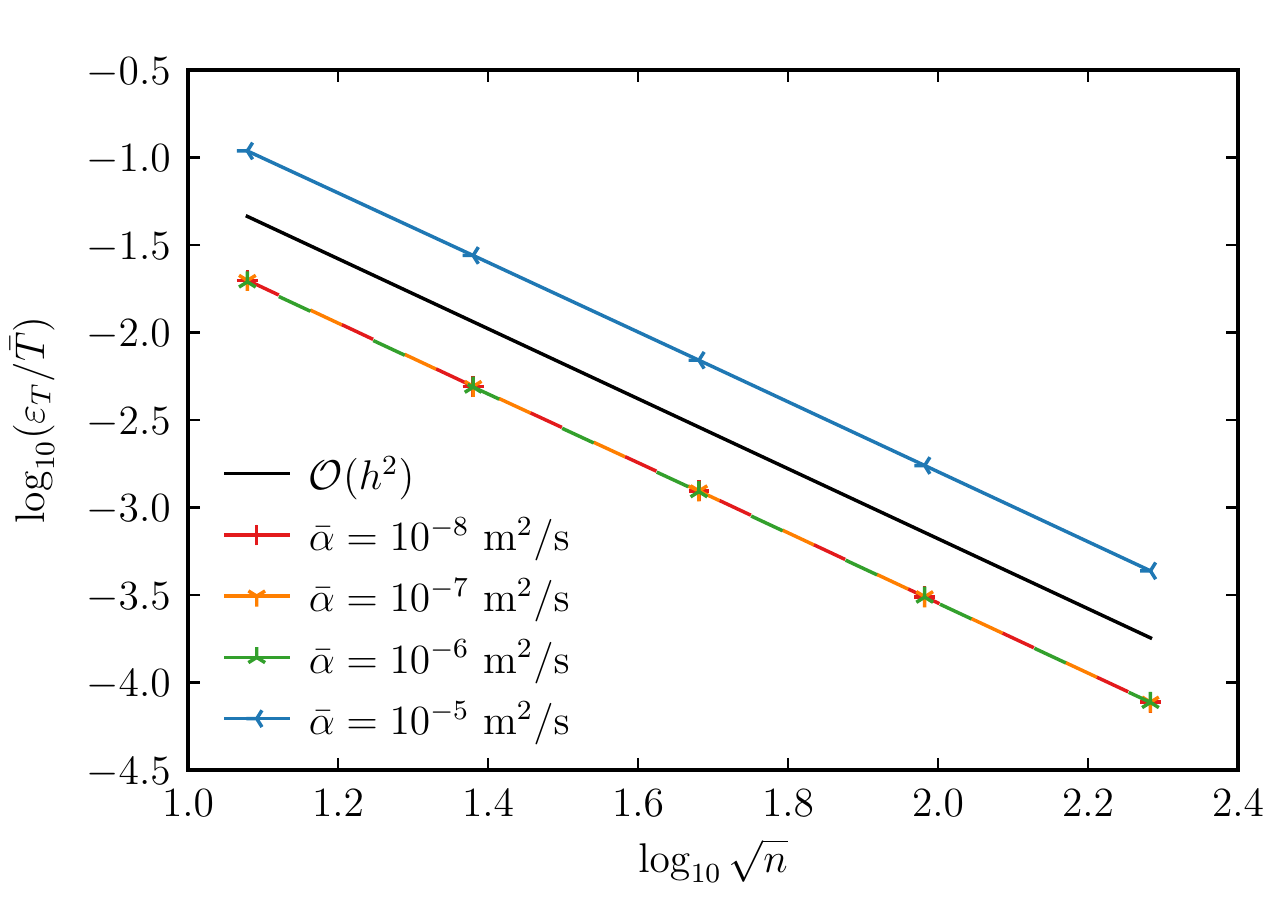}
\caption{\reviewerThree{$\epsilon\ne 0$}, $\Delta t/4$}
\label{fig:polar_T_so_rad}
\end{subfigure}
\hspace{0.25em}
\begin{subfigure}[t]{.49\textwidth}
\includegraphics[scale=.64,clip=true,trim=2.25in 0in 2.8in 0in]{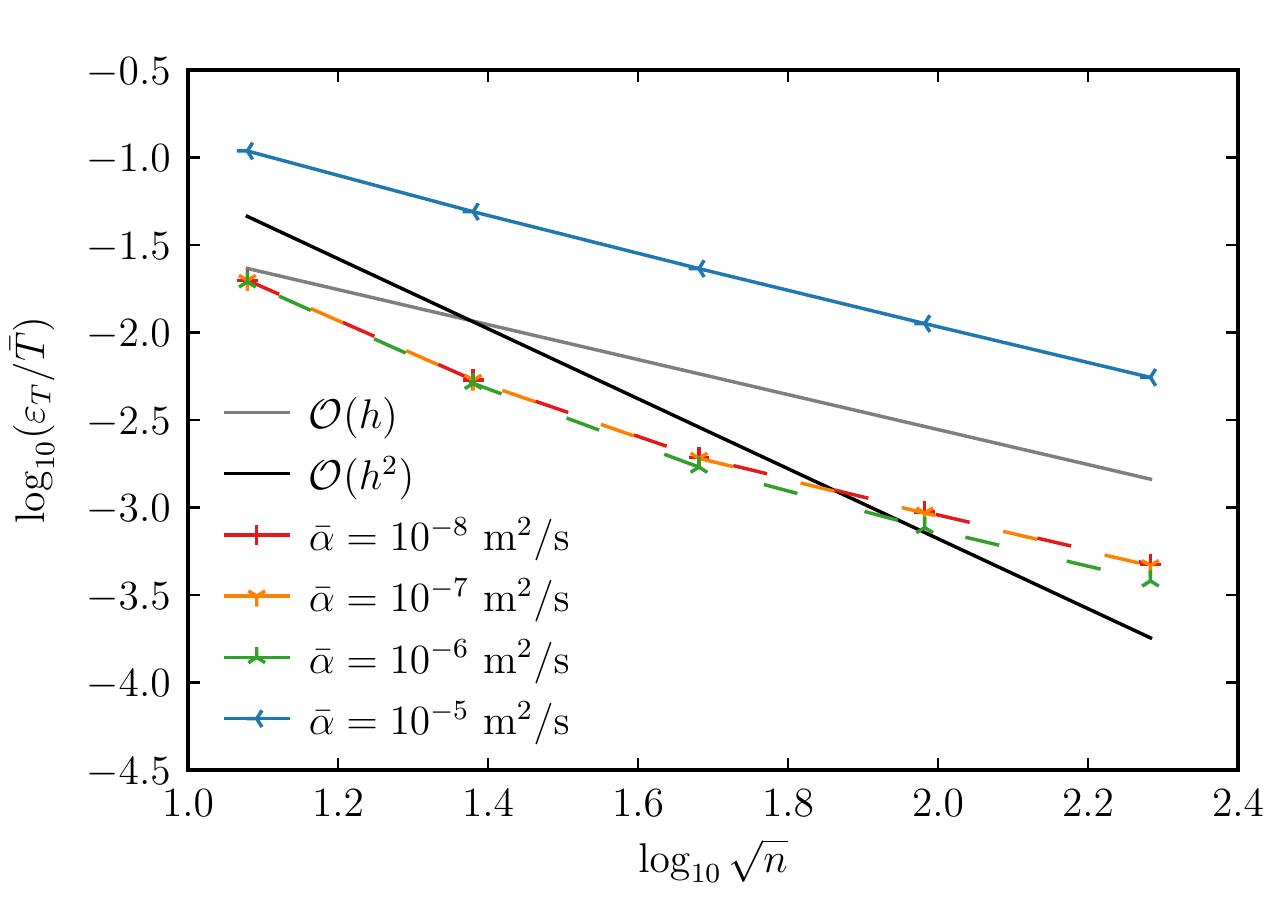}
\caption{\reviewerThree{$\epsilon\ne 0$}, \reviewerOneThree{$\Delta t/2$}}
\label{fig:polar_T_fo_rad}
\end{subfigure}
\\
\begin{subfigure}[t]{.49\textwidth}
\includegraphics[scale=.64,clip=true,trim=2.25in 0in 2.8in 0in]{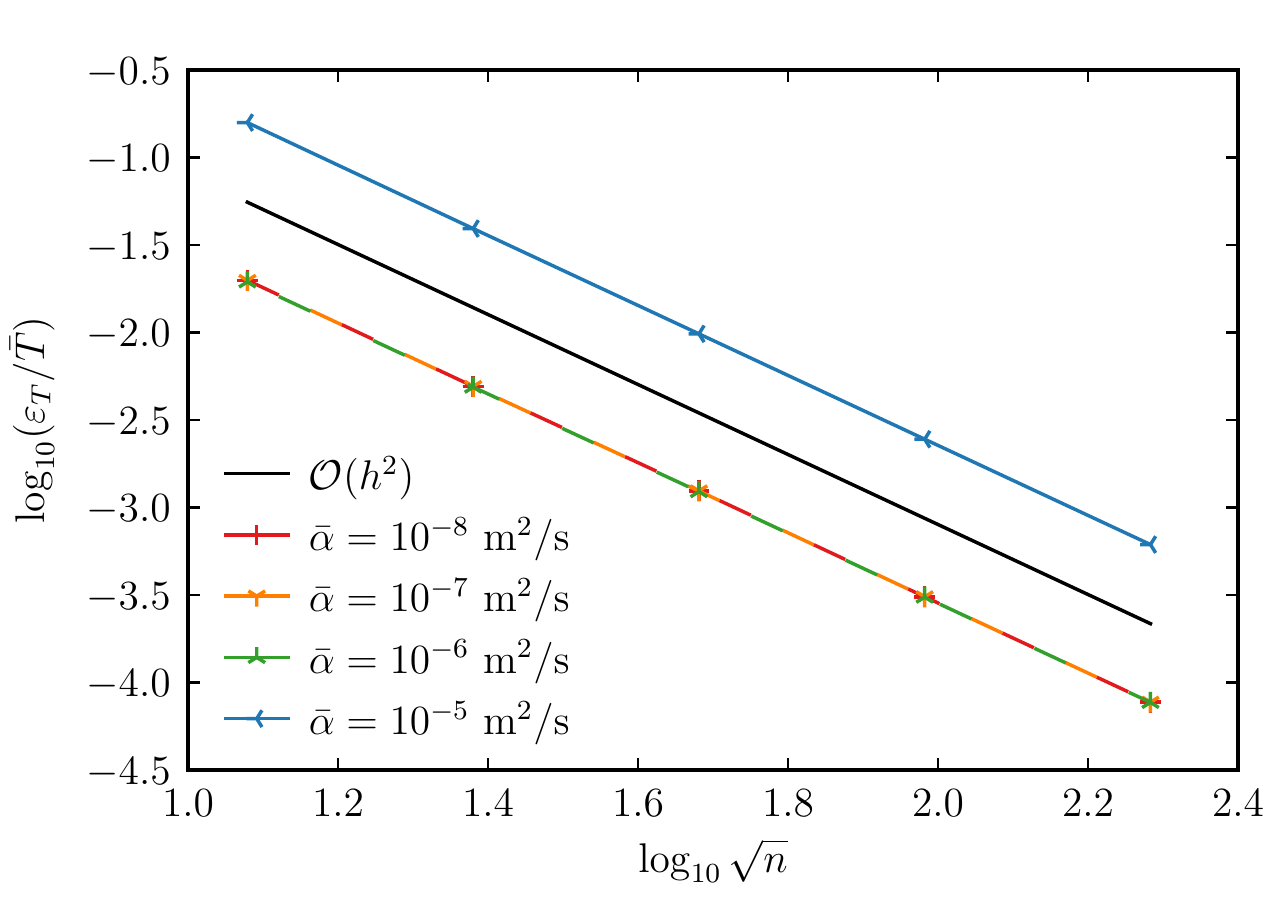}
\caption{$\epsilon=0$, $\Delta t/4$}
\label{fig:polar_T_so}
\end{subfigure}
\hspace{0.25em}
\begin{subfigure}[t]{.49\textwidth}
\includegraphics[scale=.64,clip=true,trim=2.25in 0in 2.8in 0in]{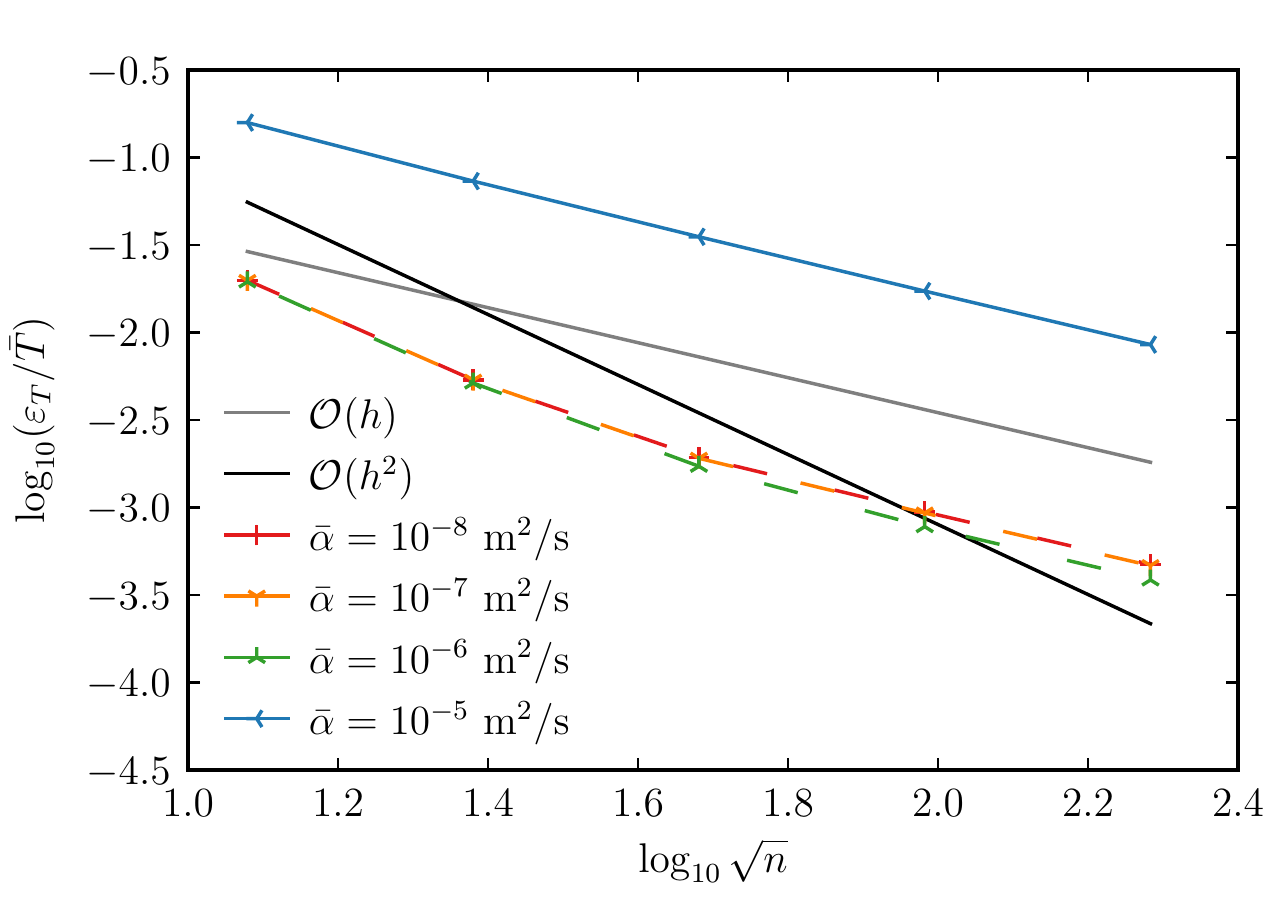}
\caption{$\epsilon=0$, \reviewerOneThree{$\Delta t/2$}}
\label{fig:polar_T_fo}
\end{subfigure}
\caption{Polar coordinates: Norm of the error for $T$.}
\vskip-\dp\strutbox
\label{fig:polar_T}
\end{figure}

\begin{figure}%[!h]
\centering
\begin{subfigure}[t]{.49\textwidth}
\includegraphics[scale=.64,clip=true,trim=2.25in 0in 2.8in 0in]{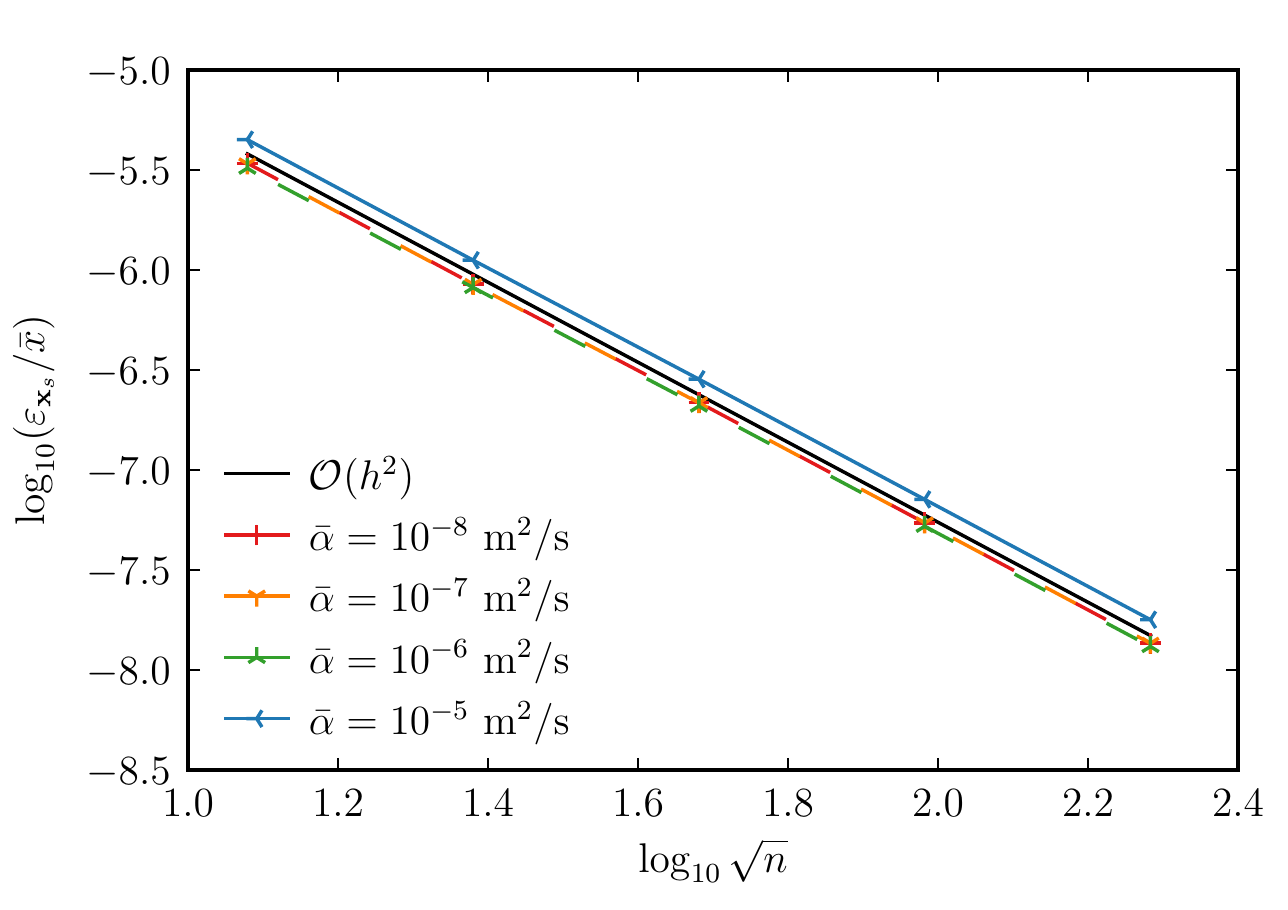}
\caption{\reviewerThree{$\epsilon\ne 0$}, $\Delta t/4$}
\label{fig:polar_xs_so_rad}
\end{subfigure}
\hspace{0.25em}
\begin{subfigure}[t]{.49\textwidth}
\includegraphics[scale=.64,clip=true,trim=2.25in 0in 2.8in 0in]{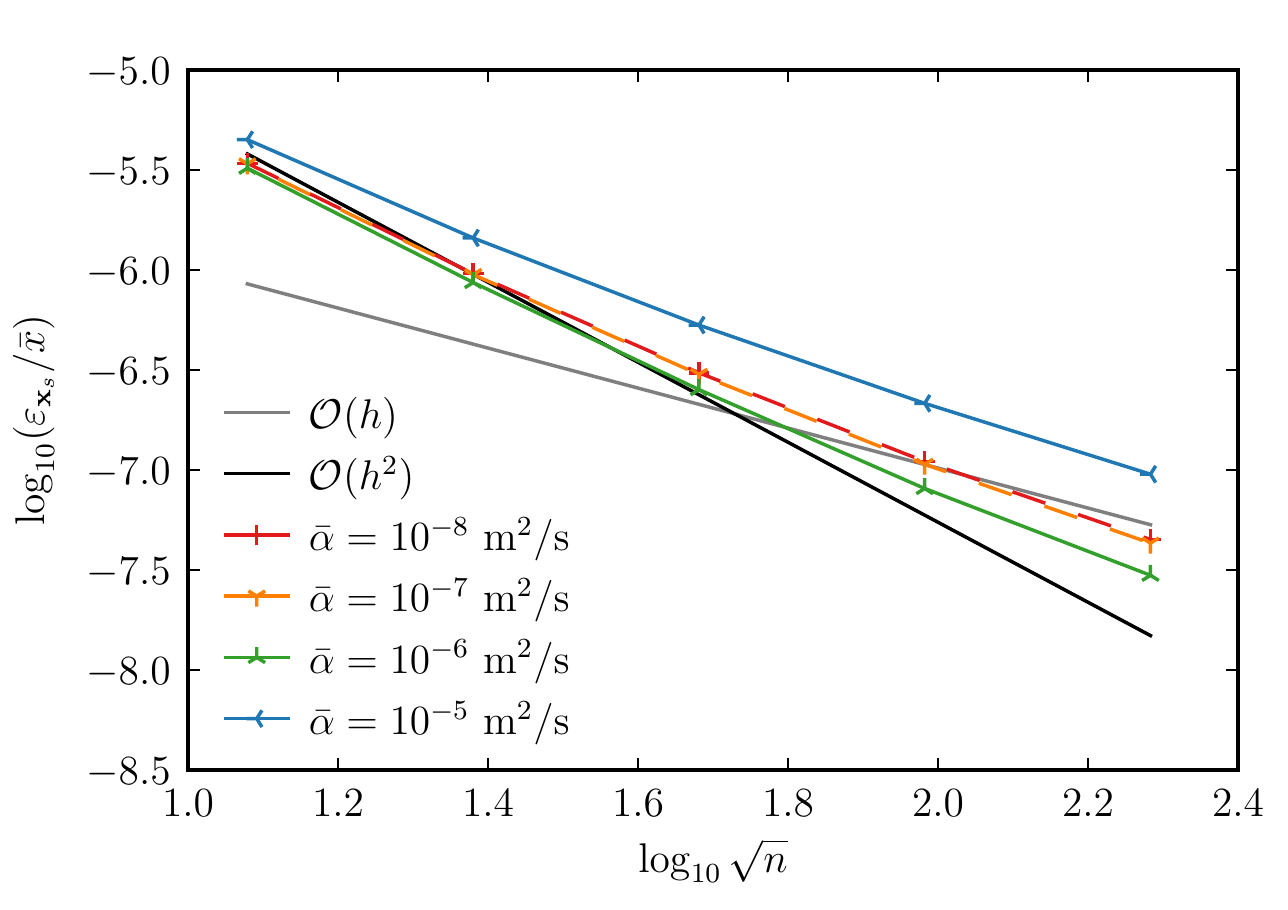}
\caption{\reviewerThree{$\epsilon\ne 0$}, \reviewerOneThree{$\Delta t/2$}}
\label{fig:polar_xs_fo_rad}
\end{subfigure}
\\
\begin{subfigure}[t]{.49\textwidth}
\includegraphics[scale=.64,clip=true,trim=2.25in 0in 2.8in 0in]{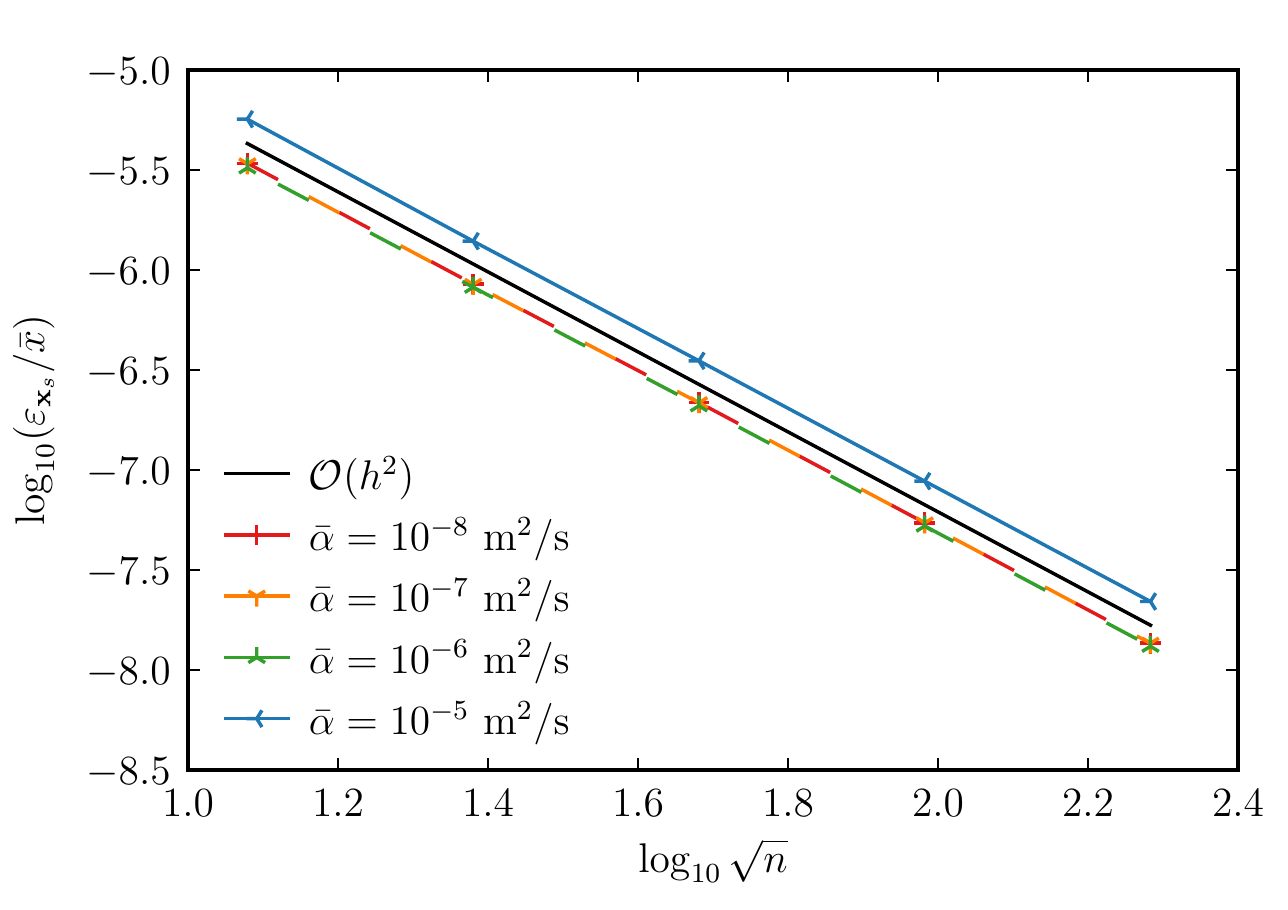}
\caption{$\epsilon=0$, $\Delta t/4$}
\label{fig:polar_xs_so}
\end{subfigure}
\hspace{0.25em}
\begin{subfigure}[t]{.49\textwidth}
\includegraphics[scale=.64,clip=true,trim=2.25in 0in 2.8in 0in]{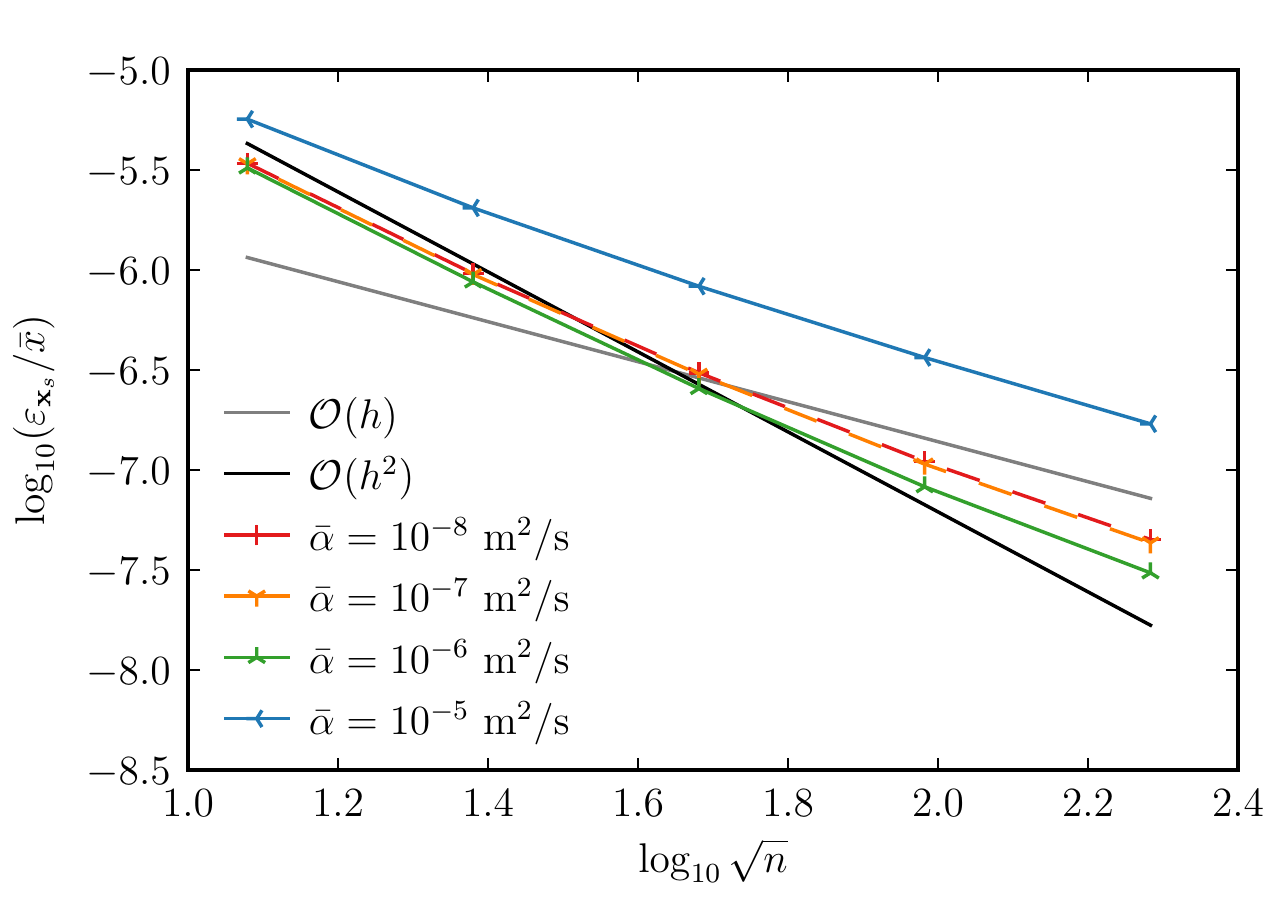}
\caption{$\epsilon=0$, \reviewerOneThree{$\Delta t/2$}}
\label{fig:polar_xs_fo}
\end{subfigure}
\caption{Polar coordinates: Norm of the error for $\mathbf{x}_s$.}
\vskip-\dp\strutbox
\label{fig:polar_xs}
\end{figure}

\reviewerOneThree{%
For each of the five values of $\bar{\alpha}$ with ($\epsilon\ne 0$) and without ($\epsilon=0$) the radiative flux, Figures~\ref{fig:polar_T} and~\ref{fig:polar_xs} show how the error norms $\varepsilon_T$~\eqref{eq:error_norm_T} and $\varepsilon_{\mathbf{x}_s}$~\eqref{eq:error_norm_xs} vary with respect to $n$.  Additionally, Tables~\ref{tab:polar_rates_T} and~\ref{tab:polar_rates_xs} provide the observed order of accuracy $p$ between discretization pairs.  The error norms in both plots and tables are $\mathcal{O}(h^2)$ for $\Delta t/4$ and approach $\mathcal{O}(h)$ for $\Delta t/2$, as expected.
}

\section{Conclusions}
\label{sec:conclusions}

In this paper, we provided an approach to perform code verification for two-dimensional, non-decomposing ablation by deriving solutions that did not require code modification.  Through this approach, we computed solutions to the heat equations for different coordinate systems, then we manufactured the dependencies of the boundary conditions.  In doing so, we could compute error norms and measure their convergence rates.  

%We demonstrated this approach for two cases, which achieved the expected accuracy.
We demonstrated this approach for two cases: one with a Cartesian coordinate system, and one with a polar coordinate system.  For both cases, we considered different thermal diffusivity values to change the relative weights between the spatial and temporal contributions to the discretization error as we assessed its convergence rate.  Both cases yielded \reviewerOneThree{the expected convergence rates given the discretization refinement ratios.}%convergence rates that were $\mathcal{O}(h^2)$.  Because the temporal discretization was refined twice as fast as the spatial discretization, these rates confirmed the expected second-order spatial accuracy and first-order temporal accuracy.
%===============================================================================
\section*{Acknowledgments} %====================================================
%===============================================================================
\label{sec:acknowledgments}
This paper describes objective technical results and analysis. Any subjective views or opinions that might be expressed in the paper do not necessarily represent the views of the U.S. Department of Energy or the United States Government.
Sandia National Laboratories is a multimission laboratory managed and operated by National Technology and Engineering Solutions of Sandia, LLC, a wholly owned subsidiary of Honeywell International, Inc., for the U.S. Department of Energy's National Nuclear Security Administration under contract DE-NA-0003525.

%%%%%%%%%%%%%%%%%%%%%%%%%%%%%%%%%%%%%%%%%%%%%%%%%%%%%%%%%%%%%%%%%%%%%%%%%%%%

\begin{table}
\centering
\reviewerOneThree{%
\begin{tabular*}{.8\textwidth}{@{\extracolsep{\fill}} c c c c c c c c c c}
\toprule
& & \multicolumn{4}{c}{$\Delta t/4$, for $\bar{\alpha}$ [m$^2$/s]} & \multicolumn{4}{c}{$\Delta t/2$, for $\bar{\alpha}$ [m$^2$/s]} \\
\cmidrule(r){3-6} \cmidrule(l){7-10}
&
Mesh & 
$10^{-8}$ &
$10^{-7}$ &
$10^{-6}$ &
$10^{-5}$ &
$10^{-8}$ &
$10^{-7}$ &
$10^{-6}$ &
$10^{-5}$ \\ 
\midrule
\multirow{4}{*}{\rotatebox[origin=c]{90}{$\epsilon\ne 0$}}
& 1--2 & 2.0136 & 2.0120 & 1.9975 & 1.9834 & 1.8909 & 1.8932 & 1.9269 & 1.1528 \\
& 2--3 & 1.9836 & 1.9829 & 1.9848 & 1.9893 & 1.4691 & 1.4813 & 1.5894 & 1.0803 \\
& 3--4 & 2.0105 & 2.0092 & 2.0017 & 1.9982 & 1.0356 & 1.0394 & 1.1482 & 1.0432 \\
& 4--5 & 2.0019 & 2.0011 & 1.9990 & 1.9993 & 0.9995 & 0.9984 & 1.0124 & 1.0220 \\
\midrule
\multirow{4}{*}{\rotatebox[origin=c]{90}{$\epsilon=0$}}
& 1--2 & 2.0136 & 2.0123 & 1.9975 & 2.0120 & 1.8908 & 1.8926 & 1.9199 & 1.1108 \\
& 2--3 & 1.9836 & 1.9829 & 1.9844 & 1.9979 & 1.4692 & 1.4820 & 1.5808 & 1.0570 \\
& 3--4 & 2.0105 & 2.0093 & 2.0021 & 2.0002 & 1.0357 & 1.0395 & 1.1459 & 1.0303 \\
& 4--5 & 2.0019 & 2.0012 & 1.9989 & 1.9999 & 0.9995 & 0.9984 & 1.0120 & 1.0154 \\
\bottomrule
\end{tabular*}
}
\caption{Polar coordinates: Observed order of accuracy $p$ for $T$.}
\label{tab:polar_rates_T}
\end{table}

%%%%%%%%%%%%%%%%%

%%%%%%%%%%%%%%%%%
\begin{table}
\centering
\reviewerOneThree{%
\begin{tabular*}{.8\textwidth}{@{\extracolsep{\fill}} c c c c c c c c c c}
\toprule
& & \multicolumn{4}{c}{$\Delta t/4$, for $\bar{\alpha}$ [m$^2$/s]} & \multicolumn{4}{c}{$\Delta t/2$, for $\bar{\alpha}$ [m$^2$/s]} \\
\cmidrule(r){3-6} \cmidrule(l){7-10}
&
Mesh & 
$10^{-8}$ &
$10^{-7}$ &
$10^{-6}$ &
$10^{-5}$ &
$10^{-8}$ &
$10^{-7}$ &
$10^{-6}$ &
$10^{-5}$ \\ 
\midrule
\multirow{4}{*}{\rotatebox[origin=c]{90}{$\epsilon\ne 0$}}
& 1--2 & 2.0061 & 2.0024 & 1.9817 & 2.0022 & 1.8325 & 1.8379 & 1.8983 & 1.6327 \\
& 2--3 & 1.9681 & 1.9655 & 1.9694 & 1.9763 & 1.6484 & 1.6581 & 1.7792 & 1.4466 \\
& 3--4 & 1.9993 & 1.9961 & 1.9964 & 1.9972 & 1.4723 & 1.4934 & 1.6427 & 1.3003 \\
& 4--5 & 1.9986 & 1.9985 & 1.9976 & 1.9982 & 1.2966 & 1.3097 & 1.4414 & 1.1767 \\
\midrule
\multirow{4}{*}{\rotatebox[origin=c]{90}{$\epsilon=0$}}
& 1--2 & 2.0062 & 2.0026 & 1.9818 & 2.0257 & 1.8325 & 1.8375 & 1.8938 & 1.4738 \\
& 2--3 & 1.9681 & 1.9655 & 1.9696 & 1.9874 & 1.6483 & 1.6572 & 1.7709 & 1.3011 \\
& 3--4 & 1.9993 & 1.9961 & 1.9964 & 1.9993 & 1.4722 & 1.4923 & 1.6313 & 1.1841 \\
& 4--5 & 1.9986 & 1.9985 & 1.9977 & 1.9989 & 1.2965 & 1.3088 & 1.4302 & 1.1011 \\
\bottomrule
\end{tabular*}
}
\caption{Polar coordinates: Observed order of accuracy $p$ for $\mathbf{x}_s$.}
\label{tab:polar_rates_xs}
\end{table}

%%%%%%%%%%%%%%%%%%%%%%%%%%%%%%%%%%%%%%%%%%%%%%%%%%%%%%%%%%%%%%%%%%%%%%%%%%%%
\clearpage

\appendix
\renewcommand{\thesection}{Appendix~\Alph{section}}
%\input{appendix_a.tex}
%\clearpage
\section{Stability Implications of $\frac{\partial q_s}{\partial T_s}$}
\label{sec:app_b}

% Heat Equation ----------------------------------------------------------------
Consider the constant-coefficient heat equation~\eqref{eq:theta_energy}.  From~\eqref{eq:bc_nonablating} and~\eqref{eq:dtheta_dn}, the boundary condition on $\Gamma_0$ is $\frac{\partial\theta}{\partial n}=0$.  From~\eqref{eq:bc_ablating} and~\eqref{eq:dtheta_dn}, the boundary condition on $\Gamma_s$ is 
\begin{align*}
\reviewerTwo{%
-\bar{k}\frac{\partial\theta}{\partial n}=-k(T_s)\frac{\partial T}{\partial n}=q_s(T_s).
}
\end{align*}

% Error ------------------------------------------------------------------------
\reviewerTwo{Consider two solutions to~\eqref{eq:theta_energy}: $\theta$ and $\tilde{\theta}$.}  From~\eqref{eq:theta_energy}, the \reviewerTwo{difference} $e_\theta = \tilde{\theta}-\theta$ is governed by
\begin{align}
\frac{\partial e_\theta}{\partial t}   - \bar{\alpha}\Delta e_\theta = 0.
\label{eq:error_equation}
\end{align}
The boundary condition on $\Gamma_0$ is $\frac{\partial e_\theta}{\partial n}=0$, and the boundary condition on $\Gamma_s$ is 
\begin{align}
\reviewerTwo{%
-\bar{k}\frac{\partial e_\theta}{\partial n}=q_s(\tilde{T}_s)-q_s(T_s).}
\label{eq:error_bc}
\end{align}
A Taylor series expansion of $q_s(\tilde{T}_s)-q_s(T_s)$ in~\eqref{eq:error_bc} about $\tilde{T}_s=T_s$ yields
\begin{align}
q_s(\tilde{T}_s)-q_s(T_s) = \frac{\partial q_s}{\partial T_s} e_{T_s} + \mathcal{O}(e_{T_s}^2),
\label{eq:error_taylor}
\end{align}
where $e_{T_s} = \tilde{T}_s - T_s$ and a Taylor series expansion of $e_{T_s}$ about $\tilde{\theta}_s=\theta_s$ yields
\begin{align}
e_{T_s} = F^{-1}(\tilde{\theta}_s) - F^{-1}(\theta_s) = \frac{1}{f(T_s)}e_{\theta_s} + \mathcal{O}(e_{\theta_s}^2).
\label{eq:error_taylor2}
\end{align}
\reviewerTwo{%
Therefore, from~\eqref{eq:error_bc},~\eqref{eq:error_taylor}, and~\eqref{eq:error_taylor2}, at $\Gamma_s$,
\begin{align}
-k(T_s)\frac{\partial e_\theta}{\partial n}= \frac{\partial q_s}{\partial T_s} e_{\theta_s} + \mathcal{O}(e_{\theta_s}^2).
\label{eq:error_taylor3}
\end{align}
}

As in Section~\ref{sec:solutions}, we express the solution to~\eqref{eq:error_equation} as 
\begin{align*}
e_\theta(\mathbf{x},t) = \reviewerThree{\sum_{i=0}^\infty\sum_{j=0}^\infty}\hat{e}_{\theta_{i,j}}(t) \varphi_{i,j}(\mathbf{x}),
%\label{eq:e_theta_form}
\end{align*}
such that
\begin{align}
\hat{e}_{\theta_{i,j}}(t) = \hat{e}_{\theta_{{i,j}_0}} e^{-\bar{\alpha}\lambda_{i,j}t}.
\label{eq:error_time}
\end{align}
If $\lambda_{i,j}$ is negative in~\eqref{eq:error_time}, $e_\theta$ will grow with time, \reviewerTwo{introducing a bifurcation}, unless $\varphi_{i,j}(\mathbf{x})=0$.

From~\eqref{eq:heat_3h}, 
\begin{align}
\Delta \varphi_{i,j}(\mathbf{x}) +\lambda_{i,j}\varphi_{i,j}(\mathbf{x})=0,
\label{eq:error_basis}
\end{align}
\reviewerTwo{%
where the boundary condition on $\Gamma_0$ is $\frac{\partial\varphi_{i,j}}{\partial n}=0$, and, from~\eqref{eq:error_taylor3}, the boundary condition on $\Gamma_s$ is
\begin{align}
-k(T_s)\frac{\partial \varphi_{i,j}}{\partial n} \approx \frac{\partial q_s}{\partial T_s}\varphi_{i,j}.
\label{eq:error_phi_bc}
\end{align}
}
Projecting~\eqref{eq:error_basis} onto $\varphi_{i,j}(\mathbf{x})$ yields
\begin{align}
\int_{\Omega(t)} \varphi_{i,j}(\mathbf{x})\Delta \varphi_{i,j}(\mathbf{x}) d\Omega +\lambda_{i,j}\int_{\Omega(t)}\varphi_{i,j}(\mathbf{x})^2 d\Omega=0.
\label{eq:error_proj}
\end{align}
Integrating the first term in~\eqref{eq:error_proj} by parts using~\eqref{eq:error_phi_bc} yields
\begin{align}
-\int_{\Gamma_s(t)} \frac{1}{k(T_s)}\frac{\partial q_s}{\partial T_s}\varphi_{i,j}(\mathbf{x_s})^2 d\Gamma-\int_{\Omega(t)} \|\nabla \varphi_{i,j}(\mathbf{x})\|^2 d\Omega +\lambda_{i,j}\int_{\Omega(t)}\varphi_{i,j}(\mathbf{x})^2 d\Omega=0.
\label{eq:error_proj2}
\end{align}
\reviewerTwo{%
In~\eqref{eq:error_proj2}, if $\varphi_{i,j}(\mathbf{x})\ne 0$, $\lambda_{i,j}$ will be positive if $\frac{\partial q_s}{\partial T_s}\ge 0$.  Therefore, if $\frac{\partial q_s}{\partial T_s}\ge 0$, $e_\theta$ will not increase in time and the bifurcation will be avoided.
}
%If $\lambda_{i,j}$ is negative and $\varphi_{i,j}(\mathbf{x})\ne 0$, \eqref{eq:error_proj2} is only satisfied if $\frac{\partial q_s}{\partial T_s}<0$.  Therefore, $\frac{\partial q_s}{\partial T_s}\ge 0$ will yield $\varphi_{i,j}(\mathbf{x})= 0$, such that $e_\theta$ will not increase in time, \reviewerTwo{and the bifurcation will be avoided}.

\addcontentsline{toc}{section}{\refname}
\bibliographystyle{elsarticle-num}

\makeatletter
\interlinepenalty=10000
\bibliography{sparc}

\begin{thebibliography}{10}
\expandafter\ifx\csname url\endcsname\relax
  \def\url#1{\texttt{#1}}\fi
\expandafter\ifx\csname urlprefix\endcsname\relax\def\urlprefix{URL }\fi
\expandafter\ifx\csname href\endcsname\relax
  \def\href#1#2{#2} \def\path#1{#1}\fi

\bibitem{roache_1998}
P.~J. Roache, Verification and Validation in Computational Science and
  Engineering, Hermosa Publishers, 1998.

\bibitem{salari_2000}
K.~Salari, P.~Knupp, Code verification by the method of manufactured solutions,
  Sandia Report SAND2000-1444, Sandia National Laboratories (Jun. 2000).
\newblock \href {https://doi.org/10.2172/759450} {\path{doi:10.2172/759450}}.

\bibitem{oberkampf_2010}
W.~L. Oberkampf, C.~J. Roy, Verification and Validation in Scientific
  Computing, Cambridge University Press, 2010.
\newblock \href {https://doi.org/10.1017/cbo9780511760396}
  {\path{doi:10.1017/cbo9780511760396}}.

\bibitem{roy_2005}
C.~J. Roy, Review of code and solution verification procedures for
  computational simulation, Journal of Computational Physics 205~(1) (2005)
  131--156.
\newblock \href {https://doi.org/10.1016/j.jcp.2004.10.036}
  {\path{doi:10.1016/j.jcp.2004.10.036}}.

\bibitem{roache_2001}
P.~J. Roache, Code verification by the method of manufactured solutions,
  Journal of Fluids Engineering 124~(1) (2001) 4--10.
\newblock \href {https://doi.org/10.1115/1.1436090}
  {\path{doi:10.1115/1.1436090}}.

\bibitem{roy_2004}
C.~J. Roy, C.~C. Nelson, T.~M. Smith, C.~C. Ober, Verification of
  {Euler/Navier}--{Stokes} codes using the method of manufactured solutions,
  International Journal for Numerical Methods in Fluids 44~(6) (2004) 599--620.
\newblock \href {https://doi.org/10.1002/fld.660} {\path{doi:10.1002/fld.660}}.

\bibitem{bond_2007}
R.~B. Bond, C.~C. Ober, P.~M. Knupp, S.~W. Bova, Manufactured solution for
  computational fluid dynamics boundary condition verification, {AIAA} Journal
  45~(9) (2007) 2224--2236.
\newblock \href {https://doi.org/10.2514/1.28099} {\path{doi:10.2514/1.28099}}.

\bibitem{veluri_2010}
S.~Veluri, C.~Roy, E.~Luke, Comprehensive code verification for an unstructured
  finite volume {CFD} code, in: 48th {AIAA} Aerospace Sciences Meeting
  Including the New Horizons Forum and Aerospace Exposition, American Institute
  of Aeronautics and Astronautics, 2010.
\newblock \href {https://doi.org/10.2514/6.2010-127}
  {\path{doi:10.2514/6.2010-127}}.

\bibitem{oliver_2012}
T.~Oliver, K.~Estacio-Hiroms, N.~Malaya, G.~Carey, Manufactured solutions for
  the {Favre}-averaged {Navier--Stokes} equations with eddy-viscosity
  turbulence models, in: 50th {AIAA} Aerospace Sciences Meeting including the
  New Horizons Forum and Aerospace Exposition, American Institute of
  Aeronautics and Astronautics, 2012.
\newblock \href {https://doi.org/10.2514/6.2012-80}
  {\path{doi:10.2514/6.2012-80}}.

\bibitem{eca_2016}
L.~E\c{c}a, C.~M. Klaij, G.~Vaz, M.~Hoekstra, F.~Pereira, On code verification
  of {RANS} solvers, Journal of Computational Physics 310 (2016) 418--439.
\newblock \href {https://doi.org/10.1016/j.jcp.2016.01.002}
  {\path{doi:10.1016/j.jcp.2016.01.002}}.

\bibitem{freno_2021}
B.~A. Freno, B.~R. Carnes, V.~G. Weirs, Code-verification techniques for
  hypersonic reacting flows in thermochemical nonequilibrium, Journal of
  Computational Physics 425 (2021).
\newblock \href {https://doi.org/10.1016/j.jcp.2020.109752}
  {\path{doi:10.1016/j.jcp.2020.109752}}.

\bibitem{chamberland_2010}
{\'E}.~Chamberland, A.~Fortin, M.~Fortin, Comparison of the performance of some
  finite element discretizations for large deformation elasticity problems,
  Computers \& Structures 88~(11) (2010) 664 -- 673.
\newblock \href {https://doi.org/10.1016/j.compstruc.2010.02.007}
  {\path{doi:10.1016/j.compstruc.2010.02.007}}.

\bibitem{etienne_2012}
S.~{\'E}tienne, A.~Garon, D.~Pelletier, Some manufactured solutions for
  verification of fluid--structure interaction codes, Computers \& Structures
  106--107 (2012) 56--67.
\newblock \href {https://doi.org/10.1016/j.compstruc.2012.04.006}
  {\path{doi:10.1016/j.compstruc.2012.04.006}}.

\bibitem{veeraragavan_2016}
A.~Veeraragavan, J.~Beri, R.~J. Gollan, Use of the method of manufactured
  solutions for the verification of conjugate heat transfer solvers, Journal of
  Computational Physics 307 (2016) 308--320.
\newblock \href {https://doi.org/10.1016/j.jcp.2015.12.004}
  {\path{doi:10.1016/j.jcp.2015.12.004}}.

\bibitem{brady_2012}
P.~T. Brady, M.~Herrmann, J.~M. Lopez, Code verification for finite volume
  multiphase scalar equations using the method of manufactured solutions,
  Journal of Computational Physics 231~(7) (2012) 2924--2944.
\newblock \href {https://doi.org/10.1016/j.jcp.2011.12.040}
  {\path{doi:10.1016/j.jcp.2011.12.040}}.

\bibitem{lovato_2021}
S.~Lovato, S.~L. Toxopeus, J.~W. Settels, G.~H. Keetels, G.~Vaz, Code
  verification of non-{Newtonian} fluid solvers for single- and two-phase
  laminar flows, Journal of Verification, Validation and Uncertainty
  Quantification 6~(2) (2021).
\newblock \href {https://doi.org/10.1115/1.4050131}
  {\path{doi:10.1115/1.4050131}}.

\bibitem{mcclarren_2008}
R.~G. McClarren, R.~B. Lowrie, Manufactured solutions for the $p_1$
  radiation-hydrodynamics equations, Journal of Quantitative Spectroscopy and
  Radiative Transfer 109~(15) (2008) 2590--2602.
\newblock \href {https://doi.org/10.1016/j.jqsrt.2008.06.003}
  {\path{doi:10.1016/j.jqsrt.2008.06.003}}.

\bibitem{ellis_2009}
J.~R. Ellis, C.~D. Hall, Model development and code verification for simulation
  of electrodynamic tether system, Journal of Guidance, Control, and Dynamics
  32~(6) (2009) 1713--1722.
\newblock \href {https://doi.org/10.2514/1.44638} {\path{doi:10.2514/1.44638}}.

\bibitem{marchand_2013}
R.~G. Marchand, The method of manufactured solutions for the verification of
  computational electromagnetic codes, phdthesis, Stellenbosch (Mar. 2013).

\bibitem{freno_em_mms_2020}
B.~A. Freno, N.~R. Matula, W.~A. Johnson, Manufactured solutions for the
  method-of-moments implementation of the electric-field integral equation,
  Journal of Computational Physics 443 (2021).
\newblock \href {https://doi.org/10.1016/j.jcp.2021.110538}
  {\path{doi:10.1016/j.jcp.2021.110538}}.

\bibitem{freno_em_mms_quad_2021}
B.~A. Freno, N.~R. Matula, J.~I. Owen, W.~A. Johnson, Code-verification
  techniques for the method-of-moments implementation of the electric-field
  integral equation, Journal of Computational Physics 451 (2022).
\newblock \href {https://doi.org/10.1016/j.jcp.2021.110891}
  {\path{doi:10.1016/j.jcp.2021.110891}}.

\bibitem{hogan_1994}
R.~Hogan, B.~Blackwell, R.~Cochran, Numerical solution of two-dimensional
  ablation problems using the finite control volume method with unstructured
  grids, in: 6th {AIAA/ASME} Joint Thermophysics and Heat Transfer Conference,
  no. AIAA 1994-2085, American Institute of Aeronautics and Astronautics, 1994.
\newblock \href {https://doi.org/10.2514/6.1994-2085}
  {\path{doi:10.2514/6.1994-2085}}.

\bibitem{blackwell_1994}
B.~F. Blackwell, R.~E. Hogan, One-dimensional ablation using {Landau}
  transformation and finite control volume procedure, Journal of Thermophysics
  and Heat Transfer 8~(2) (1994) 282--287.
\newblock \href {https://doi.org/10.2514/3.535} {\path{doi:10.2514/3.535}}.

\bibitem{amar_2008}
A.~J. Amar, B.~F. Blackwell, J.~R. Edwards, One-dimensional ablation using a
  full {Newton's} method and finite control volume procedure, Journal of
  Thermophysics and Heat Transfer 22~(1) (2008) 71--82.
\newblock \href {https://doi.org/10.2514/1.29610} {\path{doi:10.2514/1.29610}}.

\bibitem{amar_2009}
A.~J. Amar, B.~F. Blackwell, J.~R. Edwards, Development and verification of a
  one-dimensional ablation code including pyrolysis gas flow, Journal of
  Thermophysics and Heat Transfer 23~(1) (2009) 59--71.
\newblock \href {https://doi.org/10.2514/1.36882} {\path{doi:10.2514/1.36882}}.

\bibitem{amar_2011}
A.~Amar, N.~Calvert, B.~Kirk, Development and verification of the charring
  ablating thermal protection implicit system solver, in: 49th AIAA Aerospace
  Sciences Meeting including the New Horizons Forum and Aerospace Exposition,
  2011.
\newblock \href {https://doi.org/10.2514/6.2011-144}
  {\path{doi:10.2514/6.2011-144}}.

\bibitem{freno_ablation}
B.~A. Freno, B.~R. Carnes, N.~R. Matula, Nonintrusive manufactured solutions
  for ablation, Physics of Fluids 33~(1) (2021).
\newblock \href {https://doi.org/10.1063/5.0037245}
  {\path{doi:10.1063/5.0037245}}.

\bibitem{abramowitz_1964}
M.~Abramowitz, I.~A. Stegun, Handbook of mathematical functions with formulas,
  graphs, and mathematical tables, Vol.~55, United States Department of
  Commerce, National Bureau of Standards, 1964.

\bibitem{aria}
{{SIERRA} Thermal/Fluid Development Team}, {SIERRA} multimechanics module: Aria
  user manual -- version 4.56, Sandia Report SAND2020-4000, Sandia National
  Laboratories (Apr. 2020).
\newblock \href {https://doi.org/10.2172/1615880} {\path{doi:10.2172/1615880}}.

\bibitem{gent_1996}
A.~N. Gent, A new constitutive relation for rubber, Rubber Chemistry and
  Technology 69~(1) (1996).
\newblock \href {https://doi.org/10.5254/1.3538357}
  {\path{doi:10.5254/1.3538357}}.

\end{thebibliography}
\makeatother

\end{document}